\documentclass[draft]{agujournal2019}
\usepackage{setspace}
\usepackage{url} 
\usepackage{amssymb}
\usepackage{subfigure}
\usepackage{subcaption}
\usepackage[inline]{trackchanges}
\usepackage{soul}
\usepackage{enumitem}
\usepackage{adjustbox}

\draftfalse

\journalname{Geophysical Research Letters}

\begin{document}

\title{Energy partitioning between thermal and non-thermal electrons and ions in magnetotail reconnection}

\authors{Abhishek Rajhans\affil{1}, Mitsuo Oka\affil{1}, Marit Øieroset\affil{1}, Tai Phan\affil{1}, Ian J. Cohen\affil{2}, Stephen A. Fuselier\affil{3,4}, Drew L. Turner\affil{2}, James L. Burch\affil{3}, Christopher T. Russell\affil{5}, Christine Gabrielse\affil{6}, Daniel J. Gershman\affil{7}, Roy B. Torbert\affil{8}}

\affiliation{1}{University of California, Berkeley, Space Sciences Lab, CA 94720, USA}
\affiliation{2}{Space Exploration Sector, Johns Hopkins Applied Physics Laboratory, Laurel, MD 20723-6099, USA}
\affiliation{3}{Southwest Research Institute, San Antonio, TX 78238, USA}
\affiliation{4}{Department of Physics \& Astronomy, University of Texas at San Antonio, San Antonio, TX 78249, USA}
\affiliation{5}{University of California, Los Angeles, Los Angeles, CA 90095, USA}
\affiliation{6}{The Aerospace Corporation, 
El Segundo, CA 90245-4609, USA}
\affiliation{7}{NASA Goddard Space Flight Center, Greenbelt, MD 20771, USA}
\affiliation{8}{University of New Hampshire, Durham, NH 03824, USA}

\correspondingauthor{Abhishek Rajhans}{arajhans@berkeley.edu}

\justify
\begin{keypoints}

\item  Despite having softer energy spectra (power-law index $\kappa \gtrsim$ 5), ions carry more non-thermal energy density than electrons ($\kappa$ of 4-6). 
\newline

\item The non-thermal energy fraction for both ions (10-50\%) and electrons (30-60\%) rarely exceeds 50\%. \newline 

\item Both ions and electrons tend to have softer spectra for higher values of average energy of particles. \newline

\end{keypoints}

\begin{abstract}
Magnetic reconnection is an explosive energy release event. It plays an important role in accelerating particles to high non-thermal energies. These particles often exhibit energy spectra characterized by a power-law distribution. However, the partitioning of energy between thermal and non-thermal components, and between ions and electrons, remains unclear. This study provides estimates of energy partition based on a statistical analysis of magnetic reconnection events in Earth's magnetotail using data from the Magnetospheric Multiscale (MMS) mission. Ions are up to ten times more energetic than electrons but have softer spectra. We found for both ions and electrons that, as the average energy of particles (temperature) increases, their energy spectra become \textit{softer} (steeper) and thus, the fraction of energy carried by the non-thermal components decreases. These results challenge existing theories of particle acceleration through magnetotail reconnection. 

\end{abstract}

\section*{Plain Language Summary}
Magnetic reconnection events are explosive processes that energize plasma and are accompanied by high-speed plasma flows. During these events, both electrons and ions are heated and accelerated, resulting in thermal and non-thermal components, respectively. We conducted a statistical analysis of reconnection events in Earth’s magnetotail to explore how energy is distributed between electrons and protons, and between their thermal and non-thermal components. Our findings show that as the average energy of particles (temperature) increases, the fraction of energy in the non-thermal particles decreases. We also observed that this fraction is lower for ions, even though they are up to ten times more energetic than electrons. We discuss the implications of these results.

\section{Introduction}\label{sec:introduction}
Magnetic reconnection is a process where the energy stored in magnetic fields is released explosively and converted into kinetic energy of particles. This process results in increase in plasma temperature (heating), acceleration of particle to non-thermal energies, and bi-directional jets (e.g. \citeNP{priestforbes,yamadareview}). To fully understand the energetics of these events, it is essential to comprehend the energy partition between thermal and non-thermal particles. Non-thermal particles are typically characterized by a power-law form in their energy spectrum (e.g. \citeNP{pierrard2010}). Investigating this power-law tail is necessary for understanding the underlying mechanisms.

Magnetic reconnection plays a crucial role in the evolution of near-Earth space plasma (e.g. \citeNP{dungey1961}). Particularly, the magnetotail region is ideal for studying magnetic reconnection due to its geometric simplicity (e.g. \shortciteNP{nagai1998,nagai2015,oierioset2023}). In fact, particle acceleration during magnetotail reconnection events has been documented in numerous studies for both electrons and ions (e.g. \shortciteNP{oieroset2002,cohen2017jgr,cohen2021grl,okaetal2022,nagai2015,ergun2018,ergun2020}).

\citeA{christon1988,christon1989,christon1991} conducted statistical studies in Earth’s central plasma sheet using analytical models for fitting energy spectra. They found that even during periods of low AE index, non-thermal particles can be energetically significant, with spectra as hard (smaller $\kappa$) as those during the periods of high AE index. The AE index represents the global geomagnetic activity level in the auroral zone and in the magnetotail. However, it is also important to study how the spectral features depend on more local parameters in the context of magnetotail reconnection.

Therefore we study energy partition between thermal and non-thermal particles, as well as between electrons and ions during Earth's magnetotail reconnection. We explore how the power-law indices which are related to the non-thermal energy fractions vary with locally measured plasma parameters (i.e. moments of distribution function of plasma, electromagnetic fields, and quantities derived from their combination ). More specifically, we examine the electron and ion energy spectra gathered during encounters of Magnetospheric Multiscale (MMS) spacecraft  with reconnection flows in the plasma sheet center over a five-year period from 2017 to 2021.

\section{Instruments and data products}\label{sec:instrumentanddata}

The Magnetospheric Multiscale (MMS) mission \cite{burchmms2016} aims to understand fundamental plasma processes in the magnetosphere, with a particular focus on studying the electron-scale physics of magnetic reconnection \cite{torbert2018science}. The mission comprises four identical spacecrafts each equipped with multiple instruments. Below, we describe only the instruments relevant to our analysis.

The Fast Plasma Investigation (FPI) \cite{fpipollock} observes both electrons and ions. The Hot Plasma Composition Analyzer (HPCA) \cite{hpcayoung} detects ions and differentiates between species such as protons, helium (He+, He++), and oxygen (O+). The Energetic Particle Detector (EPD) \cite{epdeisfeepsmauk} measures particles at higher energies (greater than 30 keV). It includes the Fly’s Eye Energetic Particle Sensor (FEEPS) for studying higher-energy electrons and ions, and the Energetic Ion Spectrometer (EIS) for examining different ion species at high energies. Additionally, we also used fast-mode data from the FIELDS instrument \cite{fieldstorbert} for electric and magnetic fields.  

For ions, our analysis centered on protons, using combined data from HPCA (survey mode) and EIS (survey mode). For electrons, we utilized combined data from FPI (fast mode) and FEEPS (survey mode). The detection limit was set using a one-count level for FEEPS and FPI, while for HPCA and EIS, we used data from the lobe region as a proxy for the detection limit. To derive an energy spectrum, we averaged the distribution functions over all directions. To reduce spurious fluctuations, we averaged the particle data over 30-second intervals. On testing with other averaging intervals (10–60 seconds) we find that the main conclusions of this paper were not affected by this choice. The plasma parameters  measured locally, were also averaged over the same time intervals for studying their systematic variation with the softness (hardness) of the energy spectra of particles. 

\section{Methodology}\label{sec:examplecases}

In this study, we only use data from instruments onboard MMS-3. To present a comprehensive overview of our methodology, we choose two example events from July 11 2017 and July 26 2017 characterized by a reversal of high-speed flows, which indicates X-line crossing. These events were originally reported by \citeA{torbert2018science} and \citeA{ergun2018}, respectively, and have been thoroughly investigated (e.g., \citeNP{genestretijgr2018,nakamura2018jgr,nakamura2019jgr,burch2019grl,okaetal2022,oierioset2023}). The  reconnection flows during the event on July 11 2017 was characterized by high density ($\sim$ 0.1 cm$^{-3}$) and low temperatures ($\sim$ 5 keV for protons and $\sim$ 2 keV for electrons). The reconnection flows during the event on July 26 2017 was marked by low density ($<$ 0.1 cm$^{-3}$) and high temperatures ($\sim$ 10 keV for protons and $\sim$ 6 keV for electrons). Additionally, significant fluctuations in magnetic fields were observed during this event.

\begin{figure}[ht]
\begin{adjustbox}{margin=-2.5cm 0cm 0cm 2.5cm} 
\includegraphics[width=1.4\textwidth]{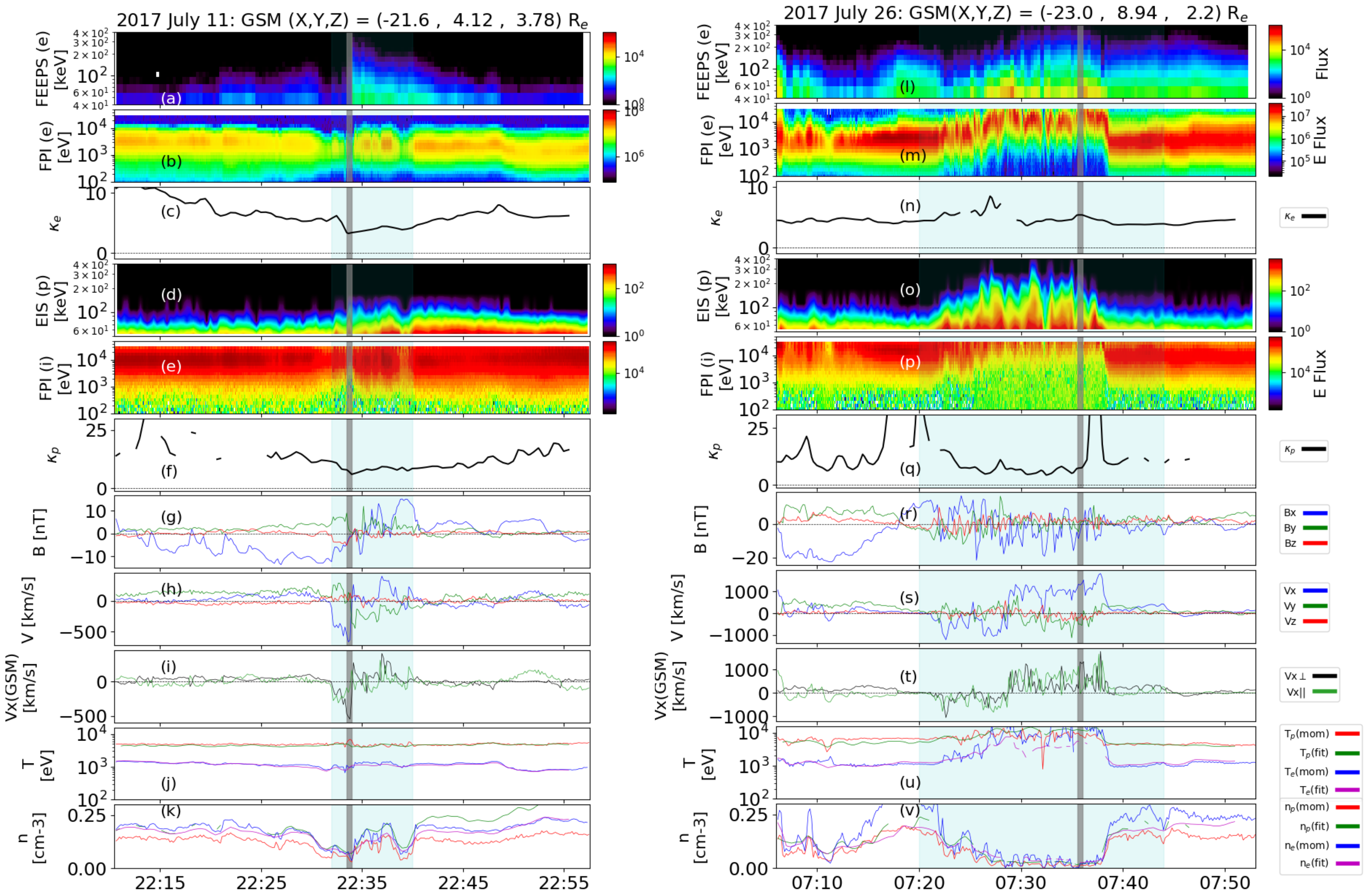}
\end{adjustbox}
\caption{The left column (panels a-k) and right column (panels l-v) correspond to two different events. Panels (a,l) and (b,m) display electron spectra from FEEPS and FPI, respectively. Panels (d,o) and (e,p) show ion spectra from EIS and FPI, respectively. Power-law indices for electrons ($\kappa_{e}$) and protons ($\kappa_{p}$) are shown in panels (c,n) and (f,q), respectively. Panels (g,r), (h,s), and (i,t)  show the magnetic fields, velocities, and components of $V_{x}$ (parallel and perpendicular to the local magnetic field), respectively (vector quantities are in GSM coordinates). Panels (j,u) and (k,v) display the temperature and density, respectively of both electrons and protons, obtained using fitting and moments data (HPCA for protons and FPI for electrons). Regions with strong flows are shaded in cyan. The grey stripes represent examples of 30 s durations during strong flows, for which the energy spectra of particles are shown in Figure~\ref{fig:distributionplots}.}\label{fig:twocases}
\end{figure}

The time series of various plasma parameters for the event on July 11 2017 are shown in Figure~\ref{fig:twocases} (a-k). The time period shaded in cyan indicates reversal of reconnection flows from tailward to earthward direction (see the blue curves in Figure~\ref{fig:twocases} h) and are accompanied by an increase in the flux of energetic particles (see Figure~\ref{fig:twocases} (a~\&~d)). The ion and electron temperatures for this event does not exceed 6 keV and 2 keV, respectively (see Figure~\ref{fig:twocases} j). The densities remain larger than 0.1 cm$^{-3}$ (see Figure~\ref{fig:twocases} k). The variations in power-law indices for electrons and protons, obtained using the procedures described later in this section, are also shown in Figure~\ref{fig:twocases} (c~\&~f). It is evident that energy spectrum is hardest for both protons and electrons during the X-line crossing.

The time series of various plasma parameters for the event on July 26 2017 are shown in Figure~\ref{fig:twocases} (l-v). Duration of reversal of reconnection flows from tailward to earthward direction are shaded in cyan (see Figure~\ref{fig:twocases} s). There is an enhancement in flux of energetic particles during this period (see Figure~\ref{fig:twocases} (l~\&~o)). Both the ion and electron temperature increases during the reconnection flow interval, reaching 10 keV and 6 keV, respectively (see Figure~\ref{fig:twocases} u). The densities are low ($< 0.1 $ cm$^{-3}$, see Figure~\ref{fig:twocases} v). Both protons and electrons show a hardening of the energy spectrum close to the X-line crossing (see Figures~\ref{fig:twocases} (n~\&~q)).

Particle energy spectra can be interpreted by different spectral models. A typical model involves combining a Maxwellian distribution (representing the thermal component) and a power-law distribution with a lower-energy cutoff $E_{c}$ (representing the non-thermal component). However, this approach requires an arbitrary choice of $E_{c}$. An alternative approach is to use the kappa model (e.g. \citeNP{olbert1968,pierrard2010}).  In this model, the spectrum transitions smoothly from a Maxwellian at lower energies to a power-law at higher energies, without any spectral break, thereby removing the need for $E_{c}$. Consequently, the spectrum is characterized by three parameters: power-law index ($\kappa$), density ($n$), and temperature ($T$). It should be emphasized that, in our study, the temperature obtained from kappa model is defined as the second order moment of the particle distribution (including both thermal and non-thermal components) in the plasma rest frame. The thermal component can be defined by embedding a Maxwellian distribution within the kappa distribution, ensuring that the particle flux matches at the most probable energy (or speed). This Maxwellian distribution effectively represents the bulk of the lower-energy population in the kappa distribution (for details, see \shortciteNP{oka2013,oka2015}). By subtracting the adjusted Maxwellian from the kappa distribution, we can determine the non-thermal fractions of particle energy ($R_{\epsilon}$) and density ($R_{n}$).

\begin{figure}[h]
\centering 
\includegraphics[width=1\textwidth]{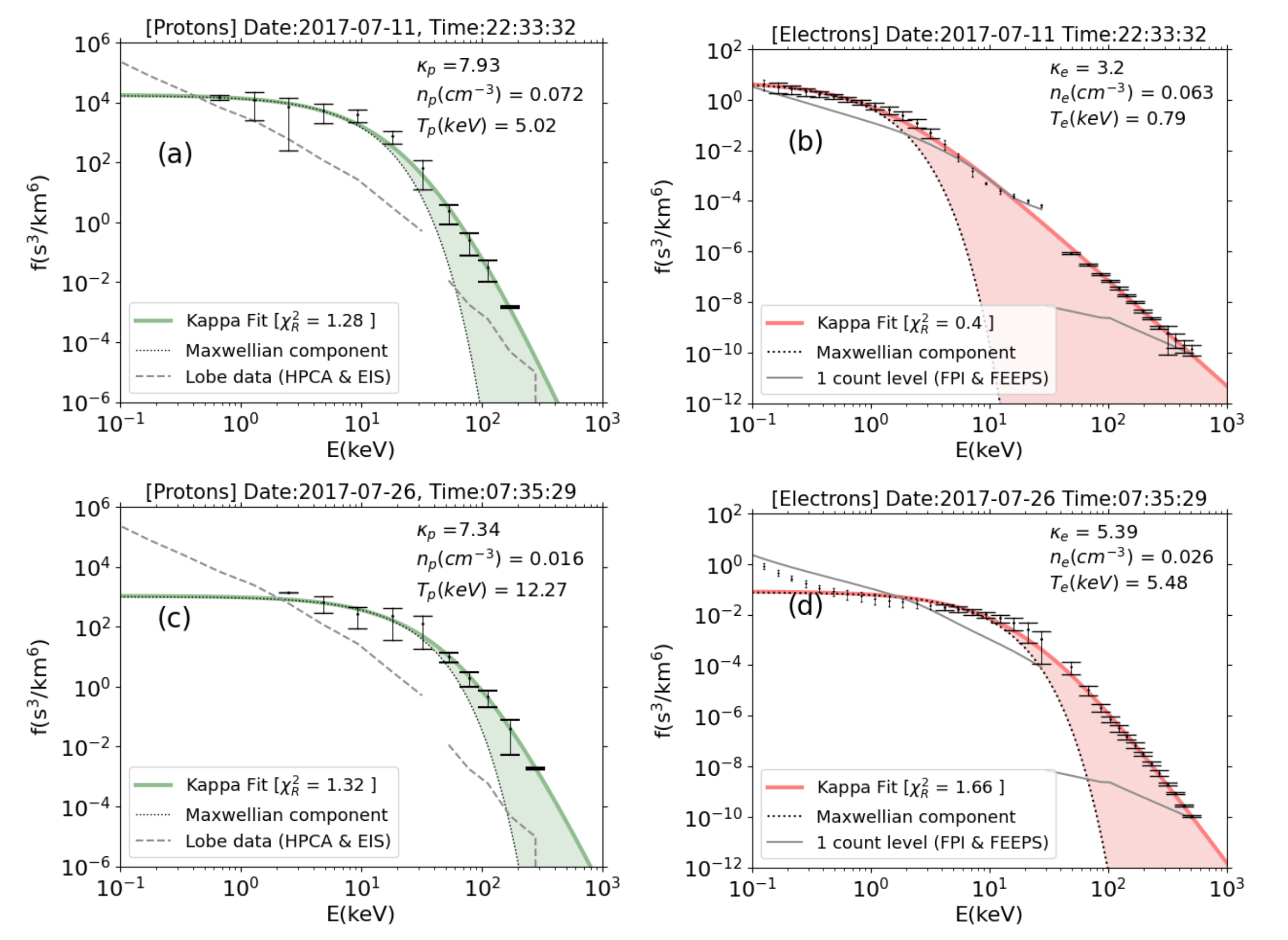}
\caption{ Panels (a \& b) show the energy spectra of protons and electrons, respectively, averaged over the time window shown in grey for event on July 11 2017 in Figure~\ref{fig:twocases}. Panels (c \& d) show the same but for event on July 27 2017. The standard deviation in measurements over 30 s interval are shown as horizontal bars on each data point. The shaded regions show the contribution of non-thermal population of particles in our definition to $f(E)$. The spectral fitting was conducted using data points that exceeded the one-count level (for FPI and FEEPS) and lobe data (for HPCA and EIS).}\label{fig:distributionplots}
\end{figure}

Figure~\ref{fig:twocases} shows our example spectral analysis. The data are taken from time periods of 30 s marked by grey stripes in Figure~\ref{fig:twocases} (a-k) and (l-v) for the events on the 11th and 27th of July 2017, respectively. These observed energy spectra for protons and electrons as well as the best fit model obtained using the kappa distribution are shown in Figure~\ref{fig:distributionplots} (a~\&~b) for the event on July 11 2017 and in panels (c~\&~d) for the event of July 26 2017. The inset text lists the fitting parameters: $\kappa, n$, and $T$. The subscripts $e$ and $p$ denote electrons and protons, respectively. The shaded regions show the contribution of non-thermal population of particles in our definition to $f(E)$.  For the event on 2017 July 11, $\kappa_{p} \sim 8$ ($R_{\epsilon,p} \sim 25\%)$ and $\kappa_{e} \sim 3$ ($R_{\epsilon,p} \sim 60\%)$. For the event on 2017 July 26, $\kappa_{p} \sim 7$ ($R_{\epsilon,p} \sim 30\%)$ and $\kappa_{e} \sim 5$ ($R_{\epsilon,p} \sim 40\%)$. 

The obtained values of $R_{\epsilon,p}$ and $R_{\epsilon,e}$ show that protons and electrons carry a significant amount of energy in non-thermal form during reconnection flows, irrespective of the plasma conditions. In both cases, the electron spectra were harder than protons ($\kappa_{e} < \kappa_{p}$). Additionally, $\kappa_{e}$ is higher at higher temperature and lower densities, whereas $\kappa_{p}$ has similar values at temperatures differing by a factor of 2 and densities differing by factor of 5. In order to check if these results are statistically significant we conducted a statistical analysis of multiple samples of reconnection flow. These results are discussed in Section~\ref{sec:statisticalanalysis}.

A caveat here is that we have used particle distribution data in the spacecraft’s frame of reference. Ideally, the analysis should be conducted in the frame moving with the bulk-flow velocity. For electrons, observed temperatures of 0.1{--}10 keV  correspond to thermal velocities of 6$\times10^{3}${--}6$\times10^{4}$ km/s, which are significantly higher than the reconnection flow velocities (4$\times10^{2}${--}1.4$\times10^{3}$ km/s) by close to an order of magnitude. Therefore the effect of shifting frames can be neglected for electrons. However, this is not the case for protons, since their thermal velocities for temperatures between 1 and 10 keV range from 1.4$\times10^{3}$ to 1.4$\times10^{4}$ km/s, which is comparable to the bulk-flow velocities. Transforming the energy spectrum from the spacecraft frame to the bulk-flow frame is challenging. At low energies ($<$1 keV), the observed energy spectrum in the spacecraft frame approaches the detection limit, making the data unreliable. Therefore, the data in these low-energy ranges need to be properly corrected and/or interpolated before performing a frame transformation. Additionally, the velocity space is binned differently in the
FEEPS and EIS energy ranges, further complicating the interpolation process.

\begin{figure}[h]
\centering
\includegraphics[width=1\textwidth]{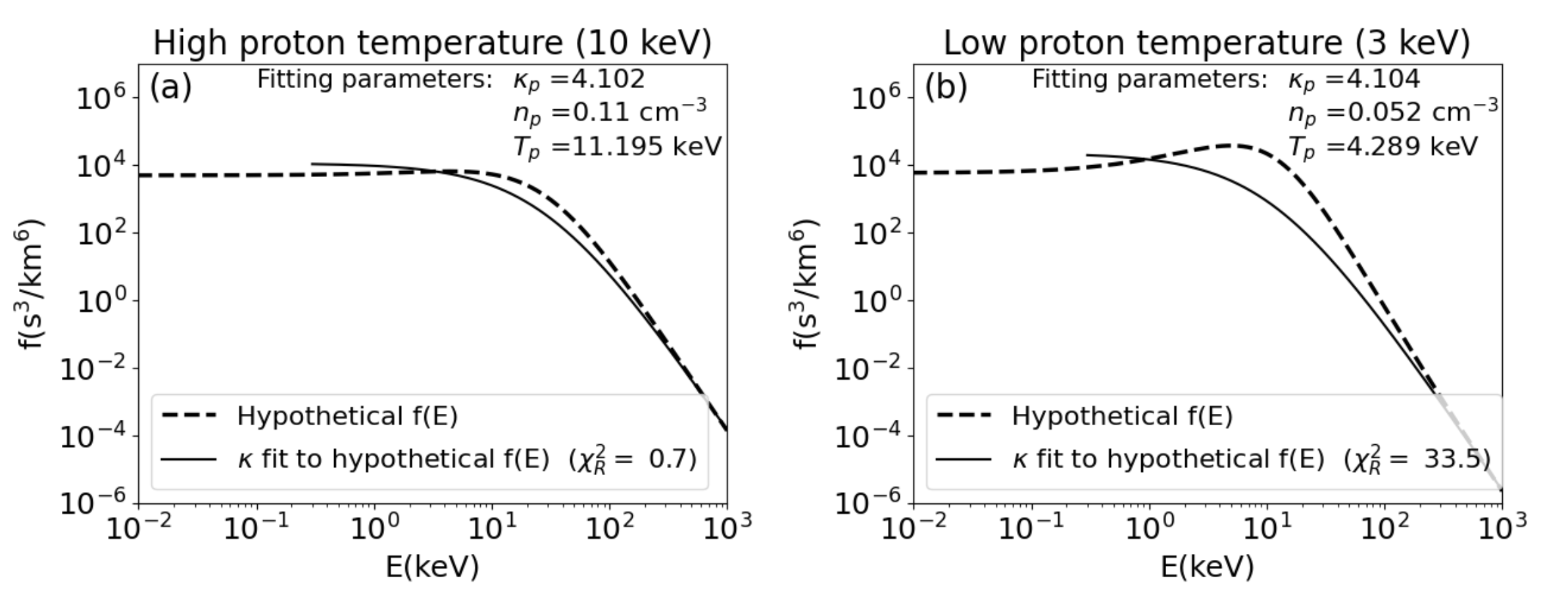}
\caption{Dashed line shows energy spectra in spacecraft frame obtained from hypothetical energy spectrum in bulk-flow frame for $\kappa = 4$, $n = 0.1$ cm$^{-3}$, and $T$ = 10 keV (for panel a) \& 3 keV (for panel b), assuming bulk-flow velocities of 1000 km/s. The solid curve shows the fit to the energy spectrum in spacecraft frame by a kappa model in the energy range of 0.1{--}1000 keV. }\label{fig:frames}

\end{figure}

To explorer the range of conditions under which a frame transformation is not required, we conducted a numerical experiment, as described below. We first calculated the theoretical velocity distribution in the bulk-flow frame ($f_{b}(v)$) for various combinations of $\kappa$, $n$, and $T$. We then transformed this velocity distribution to the spacecraft frame ($f_{s}(v)$), assuming bulk-flow velocities ranging from 400 to 1400 km/s. The theoretical velocity distributions in both frames were converted to energy spectra ($f_{b}$ and $f_{s}$). Next, we fitted $f_{s}$ within the energy range of 0.1{--}1000 keV using a different kappa model ($f_{bs}$), treating $f_{s}$ as though it was obtained in the bulk-flow frame. In all cases where the fit was satisfactory, the parameters obtained were not significantly different from those originally used to generate $f_{s}$. As expected, the quality of the fits began to deteriorate ($\chi_{R}^{2}>10$) for larger velocities ($\gtrsim 1000$ km/s). For the case of 1000 km/s, the fits started deteriorating for temperatures below 4 keV. As the velocities increased beyond 1000 km/s, the temperature below which the fit deteriorated, also increased. Figure~\ref{fig:frames} presents an example where $f_{bs}$ both successfully (panel a) and unsuccessfully (panel b) reproduced $\kappa, n$, and $T$ values close to the original parameters used to generate $f_{b}$. In MMS data, although we did see a trend of increase in temperature with bulk-flow speeds, we did not see any systematic deterioration of fit with variation in temperature and bulk flow velocity. This is due to noise in data which makes it difficult to observe peaks in energy spectra at bulk-flow speeds. However, we confirm from our analysis that irrespective of the temperature and bulk flow speeds, as long as the energy spectra observed in the spacecraft frame can be fit properly by a kappa model, the inferred parameters are not significantly in error by not accounting for transformation of the data into the bulk-flow frame.

\section{Statistical analysis}\label{sec:statisticalanalysis}

Using the methodology described in Section~\ref{sec:examplecases}, we performed a  statistical analysis. The process of sample selection consists of two steps.

In the first step we identified reconnection flows. We conducted a systematic search  for one-hour intervals where $|V_{x}| $ was $\geq$ 400 km/s in the magnetotail region (GSM $|X| > 15$ R$_{e}$ and GSM $|Y| < 10$ R$_{e}$ ). During the years 2017{--}2021, 1361 one-hour intervals meeting these criteria were found. These included time intervals encompassing events reported in previous studies \cite{torbert2018science, ergun2018, chen2019, zhou2019, huang2018, rogers2019, oierioset2023, okaetal2022, cohen2021grl}.

In the second step, we divided each event into 30-s intervals and searched for intervals where (i) $|B_{x}| \leq 10$ nT, (ii) $|V_{x}| \geq$ 300 km/s, and (iii) $V_{x\perp} \geq 0.25 |V_{x}|$ (see, e.g. \citeNP{oierioset2023}). We were left with 2321 samples. Note that, due to these requirements, most (962 out of 1361) of the one-hour durations did not contribute any sample to our study. Most of the one-hour intervals (340 out of 1361) that contributed had fewer than 10 samples. Rarely did any interval contribute more than 10 samples (59 out of 1361). We tested various combinations of thresholds for $|B_{x}|$ (5{--}15 nT), $|V_{x}|$ (300–500 km s$^{-1}$) and $V_{x\perp}$ (0.2–0.7 $|V_{x}|$), and confirmed that our results remain consistent between these different criteria.  

Each of these 30 s intervals constitutes one sample of energy spectra for our statistical analysis. It is characterized by a combination of ($\kappa$, $n$, and $T$) for electrons and/or protons, if the kappa model fits the energy spectra reasonably well. Out of the total of 2321 samples where the spacecraft encountered magnetotail exhaust jets, we obtained good fits ($\chi_{R}^{2}<10$ assuming uncertainty of $f/2$ on measured value of $f$) for 1425 (61\%) and 1604 (69\%) samples of energy spectra of protons and electrons, respectively. Then we examined the results from our analysis separately for earthward and tailward flows, since \citeA{zhou2016jgr} have reported  a higher ratio of high-energy to low-energy electron fluxes in earthward flows. 

\begin{figure}[!h]
\centering
\includegraphics[width=1\textwidth]{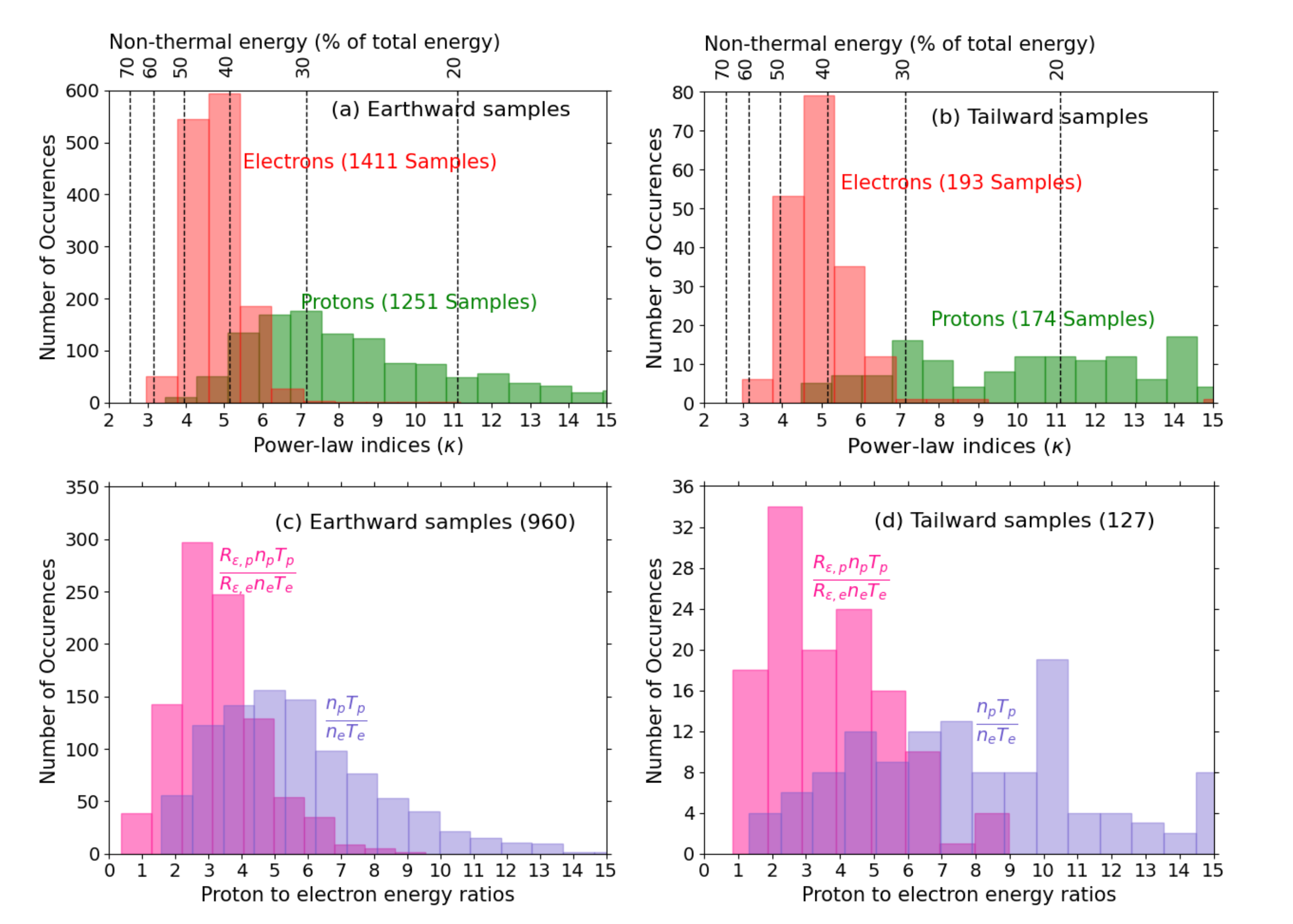}
\caption{Panel a shows the histogram of $\kappa_{e}$ (red) and $\kappa_{p}$ (green) for earthward flow samples. Vertical dashed lines show non-thermal energy fractions ($R_{\epsilon}$) for different $\kappa$. Panel c shows the histogram of ratio of non-thermal energies of protons to electrons in pink and ratio of total energies of protons to electrons in purple for earthward flow samples. Panels b and d are same as a and c, respectively, but only for tailward flow samples.}\label{fig:histogramsenergetics}
\end{figure}

Earthward flows constituted 1251 (1411) samples of energy spectra out of total 1425 (1604) samples for protons (electrons).  Tailward flows constituted 174 (193) samples of energy spectra of protons (electrons). Figure~\ref{fig:histogramsenergetics} (a) displays the histogram of power-law indices ($\kappa$) for electrons and protons obtained from all good fits in the exhaust region with earthward flows ($\sim$ 85 \% of our samples). The distribution of power-law indices in tailward flow samples ($\sim 15\%$ of the sample) are shown in Figure~\ref{fig:histogramsenergetics} (b). The distribution of power-law indices for electrons ($\kappa_{e}$) is qualitatively similar in panels a and b of Figure~\ref{fig:histogramsenergetics} and falls within a narrow region of 3{--}7 with a peak around 4{--}5. Protons, on the other hand, exhibit a much wider distribution of power-law indices ($\kappa_{p}$ $\gtrsim 4$) in both of these panels. Additionally, in Figure~\ref{fig:histogramsenergetics} (a) distribution of $\kappa_{p}$ has a peak around 7, but it exhibits a more uniform distribution in Figure~\ref{fig:histogramsenergetics} (b). There is a possibility that protons in earthward flows tend to show a harder spectrum, however, given the small sample size of tailward flows, more analysis is needed.

The values for $\kappa_{p}$ and $\kappa_{e}$ obtained in our study are consistent with \citeA{christon1991}. In terms of non-thermal energy fraction ($R_{\epsilon}$), we find that non-thermal electrons consistently carry 30{--}60\% of the total electron energy in our samples, while non-thermal protons carry 15{--}50\% of the total proton energy. Cases where the non-thermal energy fraction of either protons or electrons surpasses 50\% are uncommon (4 out of 1425 samples for protons and 102 out of 1604 samples for electrons).

Figure~\ref{fig:histogramsenergetics} (c)  illustrates the distribution of ratio of energies of non-thermal protons ($R_{\epsilon,p}n_{p}T_{p}$) and electrons ($R_{\epsilon,e}n_{e}T_{e}$) as well as the ratio of total energies of protons ($n_{p}T_{p}$) and electrons ($n_{e}T_{e}$) for earthward flow samples. The non-thermal population of protons carry up to 5 times more energy than non-thermal population of electrons, while the total energy of protons is 2{--}8  times that of electrons. Figure~\ref{fig:histogramsenergetics} (d)  illustrates these ratios for tailward flow samples. The partition of total energy and non-thermal energy between protons and electrons is similar for both earthward and tailward flow samples.

\begin{figure}[!ht]
\centering
\includegraphics[width=1\textwidth]{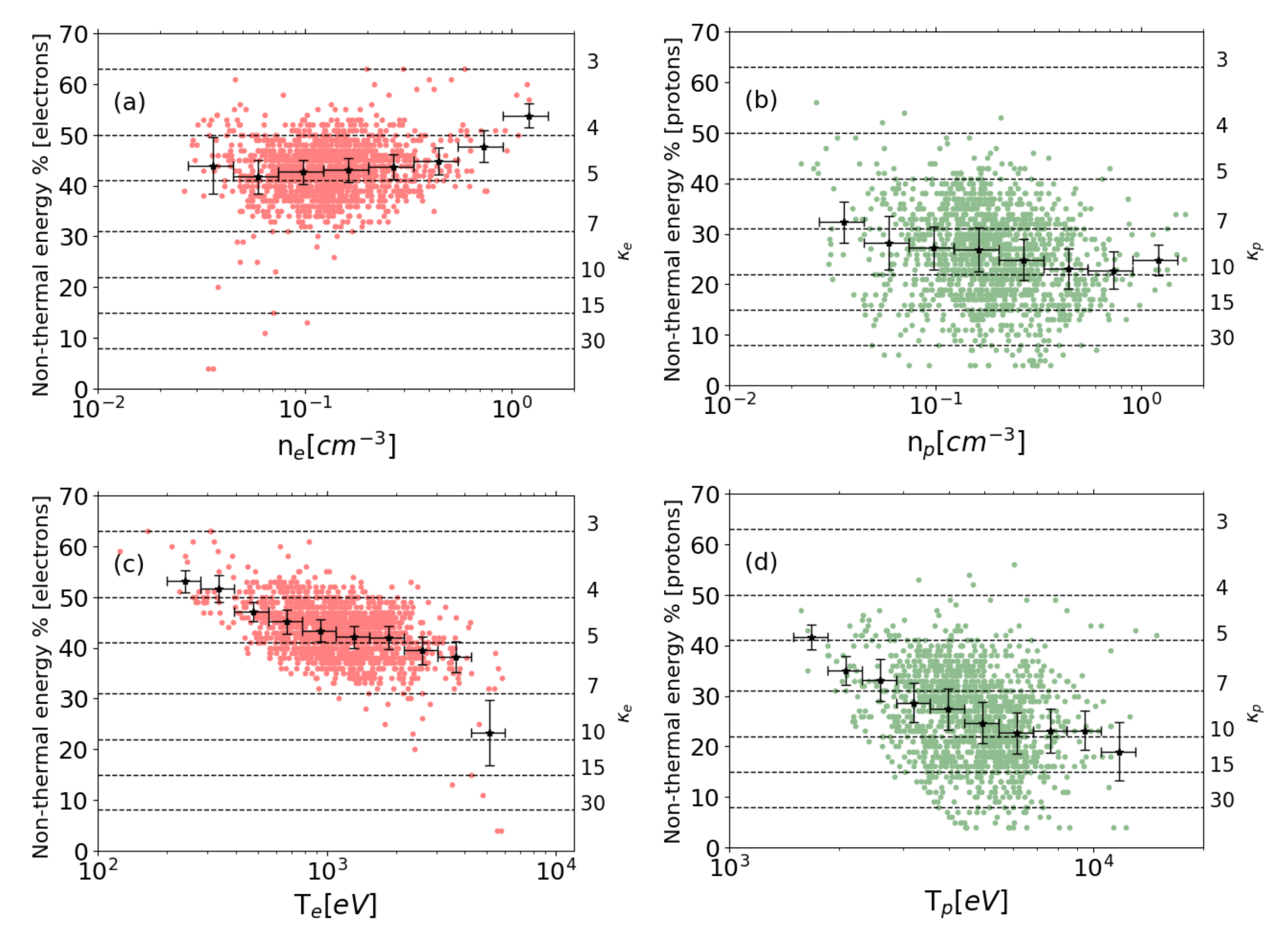}
\caption{Dependence of non-thermal energy fraction ($R_{\epsilon}$), and power-law index ($\kappa$) on density ($n$) and temperature ($T$) in the exhaust region samples. Panels (a \& c) and (b \& d) show these dependences for electrons and protons, respectively.  }\label{fig:densitytemperaturedependence}
\end{figure}

We also investigated whether $\kappa_{e}$ or $\kappa_{p}$ systematically depends on the locally measured plasma quantities such as square of Alfvén speed, Alfvénic \& sonic Mach numbers, electric \& magnetic fields, and plasma beta for both electrons \& ions. The only systematic dependence $\kappa$ had, was on temperature. The dependence on density was not as evident as the dependence on temperature. Therefore, the reason for observing a dependence on local temperature but not on local plasma $\beta$ is that the unclear influence of density and magnetic fields counterbalances the variation with temperature.  Figure~\ref{fig:densitytemperaturedependence} (a~\&~c) depict how $\kappa_{e}$ varies with $n_{e}$ and $T_{e}$, respectively. Figure~\ref{fig:densitytemperaturedependence} (b~\&~d)  illustrate the variation of $\kappa_{p}$ with $n_{p}$ and $T_{p}$, respectively. Electrons exhibit a higher non-thermal energy fraction ($R_{\epsilon,e}$) (indicated by lower $\kappa_{e}$ values i.e. harder energy spectra) at lower values of $T_{e}$ (see Figure~\ref{fig:densitytemperaturedependence} (c)). Protons show a wider spread in $\kappa_{p}$ and hence $R_{\epsilon,p}$. Hence, the trend in Figure~\ref{fig:densitytemperaturedependence} (d) is not as clear as in Figure~\ref{fig:densitytemperaturedependence} (c). However, binned data average does suggest higher  $R_{\epsilon,p}$ (lower $\kappa_{p}$) at lower $T_{p}$ 

A potential limitation of our study is that some samples might have been influenced by dipolarization fronts (e.g. \citeNP{nakamura2002}). However, we did not observe any systematic dependence of the hardness or softness of the energy spectra on B$_{z}$. We also note that we assumed that all non-thermal particles in the outflow region originate from the inflow region, rather than from pre-existing plasma in the current sheet.  Pre-existing plasma in the inflow region would still enter the outflow region via reconnection.

\section{Summary and Discussion}
The energy released in magnetic reconnection generates waves and bulk-flows. It also heats and accelerates plasmas. In this work we studied the partitioning of energy between thermal and non-thermal populations of electrons and protons in magnetotail reconnection flow exhausts. We applied the kappa model to fit the energy spectra and analyze both thermal and non-thermal components of protons and electrons. The proton temperature in our samples were 2{--}8 times higher than the electron temperature. This is consistent with values reported in previous studies  (e.g. \shortciteNP{baumjohann1989,chen2021epp,lavraudgrl2009,slavin1985jgr}).

Thermal protons have the largest share ($\sim$55{--}65\%) of the total particle energies followed by non-thermal protons ($\sim$20{--}25\%), thermal electrons($\sim$8{--}12\%) and non-thermal electrons($\sim$6{--}9\%). This suggests that most of the energy that goes into heating and acceleration of particles may be utilized for proton heating, followed by proton acceleration, electron heating and electron acceleration.  Non-thermal electrons typically carried 30{--}60\% of the total electron energy, and non-thermal protons carried 10{--}50\% of total proton energy. However, because of ions having average energy larger than that of the electrons, the non-thermal ions carried more energy than the non-thermal electrons. Electron spectra were harder with power-law indices ranging from 3{--}7, while proton power-law indices ranged from 4{--}16. 

Interestingly, these results differ from numerical predictions that protons and electrons have spectra of similar hardness ($\kappa \sim 4$) in the case of weak guide field magnetic reconnection that is applicable to magnetotail reconnection considered here \cite{zhiyuyin2024}. In simulations performed by \shortciteNP{zhang2021} protons had harder spectrum($\kappa \sim 3.7$) than electrons($\kappa \sim 4.3$). We also note that the minimum value of $\kappa$ in our samples was $\sim$3. This is important because it suggests that the dominant contribution to the non-thermal energy density comes from the low-energy population of non-thermal particles \cite{hoshino2023apj}.

We did not find any clear dependence of the hardness (softness) of energy spectrum on any locally measured plasma quantities other than temperature. Both protons and electrons tend to show lower non-thermal energy fractions (softer spectra) at higher temperatures. These results are also consistent with earlier studies of electrons \cite{imada2011JGRA,zhou2016jgr,okaetal2022}. One possible explanation for this is the presence of an electric potential, which gives the same amount of energy to electrons (e.g. \citeNP{egedal2010grl}). This would result in a constant energy shift and, when viewed on a logarithmic scale,  a softening of the power-law spectrum.  However, this explanation applies only to electrons. Ions would be decelerated under such a potential \cite{haggerty2015grl}, contrary to our finding that both ions and electrons exhibit spectral softening with increasing temperatures (or average energies). Another possibility is that higher energy particles, especially protons have larger gyroradii and thus are more likely to escape from the acceleration site in the plasma sheet in fewer gyrations, resulting in softening of the spectrum \cite{imada2015}. Lastly, thermalization and associated scattering by waves and turbulence may have contributed to the softening of energy spectra at higher temperature. A more detailed data analysis and self-consistent simulations are required to understand the effects of these mechanisms.

\section{Open research} MMS data is publicly available at \url{https://lasp.colorado.edu/mms/sdc/public/}. PYSPEDAS \cite{pyspedaspaper}, NUMPY \cite{numpypaper}, and SCIPY \cite{scipypaper} are publicly available softwares and can be accessed at \url{https://anaconda.org/}.

\bibliography{agusample}

\begin{thebibliography}{}

\bibitem [\protect \citeauthoryear {%
{Baumjohann}%
, {Paschmann}%
\BCBL {}\ \BBA {} {Cattell}%
}{%
{Baumjohann}%
\ \protect \BOthers {.}}{%
{\protect \APACyear {1989}}%
}]{%
baumjohann1989}
\APACinsertmetastar {%
baumjohann1989}%
\begin{APACrefauthors}%
{Baumjohann}, W.%
, {Paschmann}, G.%
\BCBL {}\ \BBA {} {Cattell}, C\BPBI A.%
\end{APACrefauthors}%
\unskip\
\newblock
\APACrefYearMonthDay{1989}{{\APACmonth{06}}}{}.
\newblock
{\BBOQ}\APACrefatitle {{Average plasma properties in the central plasma sheet}} {{Average plasma properties in the central plasma sheet}}.{\BBCQ}
\newblock
\APACjournalVolNumPages{Journal of Geophysical Research}{94}{A6}{6597-6606}.
\newblock
\begin{APACrefDOI} \doi{10.1029/JA094iA06p06597} \end{APACrefDOI}
\PrintBackRefs{\CurrentBib}

\bibitem [\protect \citeauthoryear {%
{Burch}%
\ \protect \BOthers {.}}{%
{Burch}%
\ \protect \BOthers {.}}{%
{\protect \APACyear {2019}}%
}]{%
burch2019grl}
\APACinsertmetastar {%
burch2019grl}%
\begin{APACrefauthors}%
{Burch}, J\BPBI L.%
, {Dokgo}, K.%
, {Hwang}, K\BPBI J.%
, {Torbert}, R\BPBI B.%
, {Graham}, D\BPBI B.%
, {Webster}, J\BPBI M.%
\BDBL {}{Le Contel}, O.%
\end{APACrefauthors}%
\unskip\
\newblock
\APACrefYearMonthDay{2019}{{\APACmonth{04}}}{}.
\newblock
{\BBOQ}\APACrefatitle {{High-Frequency Wave Generation in Magnetotail Reconnection: Linear Dispersion Analysis}} {{High-Frequency Wave Generation in Magnetotail Reconnection: Linear Dispersion Analysis}}.{\BBCQ}
\newblock
\APACjournalVolNumPages{Geophysical Research Letters}{46}{8}{4089-4097}.
\newblock
\begin{APACrefDOI} \doi{10.1029/2019GL082471} \end{APACrefDOI}
\PrintBackRefs{\CurrentBib}

\bibitem [\protect \citeauthoryear {%
{Burch}%
, {Moore}%
, {Torbert}%
\BCBL {}\ \BBA {} {Giles}%
}{%
{Burch}%
\ \protect \BOthers {.}}{%
{\protect \APACyear {2016}}%
}]{%
burchmms2016}
\APACinsertmetastar {%
burchmms2016}%
\begin{APACrefauthors}%
{Burch}, J\BPBI L.%
, {Moore}, T\BPBI E.%
, {Torbert}, R\BPBI B.%
\BCBL {}\ \BBA {} {Giles}, B\BPBI L.%
\end{APACrefauthors}%
\unskip\
\newblock
\APACrefYearMonthDay{2016}{{\APACmonth{03}}}{}.
\newblock
{\BBOQ}\APACrefatitle {{Magnetospheric Multiscale Overview and Science Objectives}} {{Magnetospheric Multiscale Overview and Science Objectives}}.{\BBCQ}
\newblock
\APACjournalVolNumPages{Space Science Reviews}{199}{1-4}{5-21}.
\newblock
\begin{APACrefDOI} \doi{10.1007/s11214-015-0164-9} \end{APACrefDOI}
\PrintBackRefs{\CurrentBib}

\bibitem [\protect \citeauthoryear {%
C.~{Chen}%
}{%
C.~{Chen}%
}{%
{\protect \APACyear {2021}}%
}]{%
chen2021epp}
\APACinsertmetastar {%
chen2021epp}%
\begin{APACrefauthors}%
{Chen}, C.%
\end{APACrefauthors}%
\unskip\
\newblock
\APACrefYearMonthDay{2021}{{\APACmonth{07}}}{}.
\newblock
{\BBOQ}\APACrefatitle {{Preservation and variation of ion-to-electron temperature ratio in the plasma sheet in geo-magnetotail}} {{Preservation and variation of ion-to-electron temperature ratio in the plasma sheet in geo-magnetotail}}.{\BBCQ}
\newblock
\APACjournalVolNumPages{Earth and Planetary Physics}{5}{4}{337-347}.
\newblock
\begin{APACrefDOI} \doi{10.26464/epp2021035} \end{APACrefDOI}
\PrintBackRefs{\CurrentBib}

\bibitem [\protect \citeauthoryear {%
L\BPBI J.~{Chen}%
\ \protect \BOthers {.}}{%
L\BPBI J.~{Chen}%
\ \protect \BOthers {.}}{%
{\protect \APACyear {2019}}%
}]{%
chen2019}
\APACinsertmetastar {%
chen2019}%
\begin{APACrefauthors}%
{Chen}, L\BPBI J.%
, {Wang}, S.%
, {Hesse}, M.%
, {Ergun}, R\BPBI E.%
, {Moore}, T.%
, {Giles}, B.%
\BDBL {}{Lindqvist}, P\BPBI A.%
\end{APACrefauthors}%
\unskip\
\newblock
\APACrefYearMonthDay{2019}{{\APACmonth{06}}}{}.
\newblock
{\BBOQ}\APACrefatitle {{Electron Diffusion Regions in Magnetotail Reconnection Under Varying Guide Fields}} {{Electron Diffusion Regions in Magnetotail Reconnection Under Varying Guide Fields}}.{\BBCQ}
\newblock
\APACjournalVolNumPages{Geophysical Research Letters}{46}{12}{6230-6238}.
\newblock
\begin{APACrefDOI} \doi{10.1029/2019GL082393} \end{APACrefDOI}
\PrintBackRefs{\CurrentBib}

\bibitem [\protect \citeauthoryear {%
{Christon}%
\ \protect \BOthers {.}}{%
{Christon}%
\ \protect \BOthers {.}}{%
{\protect \APACyear {1988}}%
}]{%
christon1988}
\APACinsertmetastar {%
christon1988}%
\begin{APACrefauthors}%
{Christon}, S\BPBI P.%
, {Mitchell}, D\BPBI G.%
, {Williams}, D\BPBI J.%
, {Frank}, L\BPBI A.%
, {Huang}, C\BPBI Y.%
\BCBL {}\ \BBA {} {Eastman}, T\BPBI E.%
\end{APACrefauthors}%
\unskip\
\newblock
\APACrefYearMonthDay{1988}{{\APACmonth{04}}}{}.
\newblock
{\BBOQ}\APACrefatitle {{Energy spectra of plasma sheet ions and electrons from \raisebox{-0.5ex}\textasciitilde50 eV/e to \raisebox{-0.5ex}\textasciitilde1 MeV during plasma temperature transitions}} {{Energy spectra of plasma sheet ions and electrons from \raisebox{-0.5ex}\textasciitilde50 eV/e to \raisebox{-0.5ex}\textasciitilde1 MeV during plasma temperature transitions}}.{\BBCQ}
\newblock
\APACjournalVolNumPages{Journal of Geophysical Research}{93}{A4}{2562-2572}.
\newblock
\begin{APACrefDOI} \doi{10.1029/JA093iA04p02562} \end{APACrefDOI}
\PrintBackRefs{\CurrentBib}

\bibitem [\protect \citeauthoryear {%
{Christon}%
, {Williams}%
, {Mitchell}%
, {Frank}%
\BCBL {}\ \BBA {} {Huang}%
}{%
{Christon}%
\ \protect \BOthers {.}}{%
{\protect \APACyear {1989}}%
}]{%
christon1989}
\APACinsertmetastar {%
christon1989}%
\begin{APACrefauthors}%
{Christon}, S\BPBI P.%
, {Williams}, D\BPBI J.%
, {Mitchell}, D\BPBI G.%
, {Frank}, L\BPBI A.%
\BCBL {}\ \BBA {} {Huang}, C\BPBI Y.%
\end{APACrefauthors}%
\unskip\
\newblock
\APACrefYearMonthDay{1989}{{\APACmonth{10}}}{}.
\newblock
{\BBOQ}\APACrefatitle {{Spectral characteristics of plasma sheet ion and electron populations during undisturbed geomagnetic conditions}} {{Spectral characteristics of plasma sheet ion and electron populations during undisturbed geomagnetic conditions}}.{\BBCQ}
\newblock
\APACjournalVolNumPages{Journal of Geophysical Research}{94}{A10}{13409-13424}.
\newblock
\begin{APACrefDOI} \doi{10.1029/JA094iA10p13409} \end{APACrefDOI}
\PrintBackRefs{\CurrentBib}

\bibitem [\protect \citeauthoryear {%
{Christon}%
, {Williams}%
, {Mitchell}%
, {Huang}%
\BCBL {}\ \BBA {} {Frank}%
}{%
{Christon}%
\ \protect \BOthers {.}}{%
{\protect \APACyear {1991}}%
}]{%
christon1991}
\APACinsertmetastar {%
christon1991}%
\begin{APACrefauthors}%
{Christon}, S\BPBI P.%
, {Williams}, D\BPBI J.%
, {Mitchell}, D\BPBI G.%
, {Huang}, C\BPBI Y.%
\BCBL {}\ \BBA {} {Frank}, L\BPBI A.%
\end{APACrefauthors}%
\unskip\
\newblock
\APACrefYearMonthDay{1991}{{\APACmonth{01}}}{}.
\newblock
{\BBOQ}\APACrefatitle {{Spectral characteristics of plasma sheet ion and electron populations during disturbed geomagnetic conditions}} {{Spectral characteristics of plasma sheet ion and electron populations during disturbed geomagnetic conditions}}.{\BBCQ}
\newblock
\APACjournalVolNumPages{Journal of Geophysical Research}{96}{A1}{1-22}.
\newblock
\begin{APACrefDOI} \doi{10.1029/90JA01633} \end{APACrefDOI}
\PrintBackRefs{\CurrentBib}

\bibitem [\protect \citeauthoryear {%
{Cohen}%
\ \protect \BOthers {.}}{%
{Cohen}%
\ \protect \BOthers {.}}{%
{\protect \APACyear {2017}}%
}]{%
cohen2017jgr}
\APACinsertmetastar {%
cohen2017jgr}%
\begin{APACrefauthors}%
{Cohen}, I\BPBI J.%
, {Mitchell}, D\BPBI G.%
, {Kistler}, L\BPBI M.%
, {Mauk}, B\BPBI H.%
, {Anderson}, B\BPBI J.%
, {Westlake}, J\BPBI H.%
\BDBL {}{Burch}, J\BPBI L.%
\end{APACrefauthors}%
\unskip\
\newblock
\APACrefYearMonthDay{2017}{{\APACmonth{09}}}{}.
\newblock
{\BBOQ}\APACrefatitle {{Dominance of high-energy ($>$150 keV) heavy ion intensities in Earth's middle to outer magnetosphere}} {{Dominance of high-energy ($>$150 keV) heavy ion intensities in Earth's middle to outer magnetosphere}}.{\BBCQ}
\newblock
\APACjournalVolNumPages{Journal of Geophysical Research (Space Physics)}{122}{9}{9282-9293}.
\newblock
\begin{APACrefDOI} \doi{10.1002/2017JA024351} \end{APACrefDOI}
\PrintBackRefs{\CurrentBib}

\bibitem [\protect \citeauthoryear {%
{Cohen}%
\ \protect \BOthers {.}}{%
{Cohen}%
\ \protect \BOthers {.}}{%
{\protect \APACyear {2021}}%
}]{%
cohen2021grl}
\APACinsertmetastar {%
cohen2021grl}%
\begin{APACrefauthors}%
{Cohen}, I\BPBI J.%
, {Turner}, D\BPBI L.%
, {Mauk}, B\BPBI H.%
, {Bingham}, S\BPBI T.%
, {Blake}, J\BPBI B.%
, {Fennell}, J\BPBI F.%
\BCBL {}\ \BBA {} {Burch}, J\BPBI L.%
\end{APACrefauthors}%
\unskip\
\newblock
\APACrefYearMonthDay{2021}{{\APACmonth{01}}}{}.
\newblock
{\BBOQ}\APACrefatitle {{Characteristics of Energetic Electrons Near Active Magnetotail Reconnection Sites: Statistical Evidence for Local Energization}} {{Characteristics of Energetic Electrons Near Active Magnetotail Reconnection Sites: Statistical Evidence for Local Energization}}.{\BBCQ}
\newblock
\APACjournalVolNumPages{Geophysical Research Letters}{48}{1}{e90087}.
\newblock
\begin{APACrefDOI} \doi{10.1029/2020GL090087} \end{APACrefDOI}
\PrintBackRefs{\CurrentBib}

\bibitem [\protect \citeauthoryear {%
{Dungey}%
}{%
{Dungey}%
}{%
{\protect \APACyear {1961}}%
}]{%
dungey1961}
\APACinsertmetastar {%
dungey1961}%
\begin{APACrefauthors}%
{Dungey}, J\BPBI W.%
\end{APACrefauthors}%
\unskip\
\newblock
\APACrefYearMonthDay{1961}{{\APACmonth{01}}}{}.
\newblock
{\BBOQ}\APACrefatitle {{Interplanetary Magnetic Field and the Auroral Zones}} {{Interplanetary Magnetic Field and the Auroral Zones}}.{\BBCQ}
\newblock
\APACjournalVolNumPages{Physical Review Letters}{6}{2}{47-48}.
\newblock
\begin{APACrefDOI} \doi{10.1103/PhysRevLett.6.47} \end{APACrefDOI}
\PrintBackRefs{\CurrentBib}

\bibitem [\protect \citeauthoryear {%
{Egedal}%
\ \protect \BOthers {.}}{%
{Egedal}%
\ \protect \BOthers {.}}{%
{\protect \APACyear {2010}}%
}]{%
egedal2010grl}
\APACinsertmetastar {%
egedal2010grl}%
\begin{APACrefauthors}%
{Egedal}, J.%
, {L{\^e}}, A.%
, {Zhu}, Y.%
, {Daughton}, W.%
, {{\O}ieroset}, M.%
, {Phan}, T.%
\BDBL {}{Eastwood}, J\BPBI P.%
\end{APACrefauthors}%
\unskip\
\newblock
\APACrefYearMonthDay{2010}{{\APACmonth{05}}}{}.
\newblock
{\BBOQ}\APACrefatitle {{Cause of super-thermal electron heating during magnetotail reconnection}} {{Cause of super-thermal electron heating during magnetotail reconnection}}.{\BBCQ}
\newblock
\APACjournalVolNumPages{Geophysical Research Letters}{37}{10}{L10102}.
\newblock
\begin{APACrefDOI} \doi{10.1029/2010GL043487} \end{APACrefDOI}
\PrintBackRefs{\CurrentBib}

\bibitem [\protect \citeauthoryear {%
{Ergun}%
\ \protect \BOthers {.}}{%
{Ergun}%
\ \protect \BOthers {.}}{%
{\protect \APACyear {2020}}%
}]{%
ergun2020}
\APACinsertmetastar {%
ergun2020}%
\begin{APACrefauthors}%
{Ergun}, R\BPBI E.%
, {Ahmadi}, N.%
, {Kromyda}, L.%
, {Schwartz}, S\BPBI J.%
, {Chasapis}, A.%
, {Hoilijoki}, S.%
\BDBL {}{Burch}, J\BPBI L.%
\end{APACrefauthors}%
\unskip\
\newblock
\APACrefYearMonthDay{2020}{{\APACmonth{08}}}{}.
\newblock
{\BBOQ}\APACrefatitle {{Particle Acceleration in Strong Turbulence in the Earth's Magnetotail}} {{Particle Acceleration in Strong Turbulence in the Earth's Magnetotail}}.{\BBCQ}
\newblock
\APACjournalVolNumPages{Astrophysical Journal}{898}{2}{153}.
\newblock
\begin{APACrefDOI} \doi{10.3847/1538-4357/ab9ab5} \end{APACrefDOI}
\PrintBackRefs{\CurrentBib}

\bibitem [\protect \citeauthoryear {%
{Ergun}%
\ \protect \BOthers {.}}{%
{Ergun}%
\ \protect \BOthers {.}}{%
{\protect \APACyear {2018}}%
}]{%
ergun2018}
\APACinsertmetastar {%
ergun2018}%
\begin{APACrefauthors}%
{Ergun}, R\BPBI E.%
, {Goodrich}, K\BPBI A.%
, {Wilder}, F\BPBI D.%
, {Ahmadi}, N.%
, {Holmes}, J\BPBI C.%
, {Eriksson}, S.%
\BDBL {}{Vaivads}, A.%
\end{APACrefauthors}%
\unskip\
\newblock
\APACrefYearMonthDay{2018}{{\APACmonth{04}}}{}.
\newblock
{\BBOQ}\APACrefatitle {{Magnetic Reconnection, Turbulence, and Particle Acceleration: Observations in the Earth's Magnetotail}} {{Magnetic Reconnection, Turbulence, and Particle Acceleration: Observations in the Earth's Magnetotail}}.{\BBCQ}
\newblock
\APACjournalVolNumPages{Geophysical Research Letters}{45}{8}{3338-3347}.
\newblock
\begin{APACrefDOI} \doi{10.1002/2018GL076993} \end{APACrefDOI}
\PrintBackRefs{\CurrentBib}

\bibitem [\protect \citeauthoryear {%
{Genestreti}%
\ \protect \BOthers {.}}{%
{Genestreti}%
\ \protect \BOthers {.}}{%
{\protect \APACyear {2018}}%
}]{%
genestretijgr2018}
\APACinsertmetastar {%
genestretijgr2018}%
\begin{APACrefauthors}%
{Genestreti}, K\BPBI J.%
, {Nakamura}, T\BPBI K\BPBI M.%
, {Nakamura}, R.%
, {Denton}, R\BPBI E.%
, {Torbert}, R\BPBI B.%
, {Burch}, J\BPBI L.%
\BDBL {}{Russell}, C\BPBI T.%
\end{APACrefauthors}%
\unskip\
\newblock
\APACrefYearMonthDay{2018}{{\APACmonth{11}}}{}.
\newblock
{\BBOQ}\APACrefatitle {{How Accurately Can We Measure the Reconnection Rate E$_{M}$ for the MMS Diffusion Region Event of 11 July 2017?}} {{How Accurately Can We Measure the Reconnection Rate E$_{M}$ for the MMS Diffusion Region Event of 11 July 2017?}}{\BBCQ}
\newblock
\APACjournalVolNumPages{Journal of Geophysical Research (Space Physics)}{123}{11}{9130-9149}.
\newblock
\begin{APACrefDOI} \doi{10.1029/2018JA025711} \end{APACrefDOI}
\PrintBackRefs{\CurrentBib}

\bibitem [\protect \citeauthoryear {%
{Grimes}%
\ \protect \BOthers {.}}{%
{Grimes}%
\ \protect \BOthers {.}}{%
{\protect \APACyear {2022}}%
}]{%
pyspedaspaper}
\APACinsertmetastar {%
pyspedaspaper}%
\begin{APACrefauthors}%
{Grimes}, E\BPBI W.%
, {Harter}, B.%
, {Hatzigeorgiu}, N.%
, {Drozdov}, A.%
, {Lewis}, J\BPBI W.%
, {Angelopoulos}, V.%
\BDBL {}{Le Contel}, O.%
\end{APACrefauthors}%
\unskip\
\newblock
\APACrefYearMonthDay{2022}{oct}{}.
\newblock
{\BBOQ}\APACrefatitle {The Space Physics Environment Data Analysis System in Python} {The space physics environment data analysis system in python}.{\BBCQ}
\newblock
\APACjournalVolNumPages{Frontiers in Astronomy and Space Sciences}{9}{}{1020815}.
\newblock
\begin{APACrefDOI} \doi{10.3389/fspas.2022.1020815} \end{APACrefDOI}
\PrintBackRefs{\CurrentBib}

\bibitem [\protect \citeauthoryear {%
{Haggerty}%
, {Shay}%
, {Drake}%
, {Phan}%
\BCBL {}\ \BBA {} {McHugh}%
}{%
{Haggerty}%
\ \protect \BOthers {.}}{%
{\protect \APACyear {2015}}%
}]{%
haggerty2015grl}
\APACinsertmetastar {%
haggerty2015grl}%
\begin{APACrefauthors}%
{Haggerty}, C\BPBI C.%
, {Shay}, M\BPBI A.%
, {Drake}, J\BPBI F.%
, {Phan}, T\BPBI D.%
\BCBL {}\ \BBA {} {McHugh}, C\BPBI T.%
\end{APACrefauthors}%
\unskip\
\newblock
\APACrefYearMonthDay{2015}{{\APACmonth{11}}}{}.
\newblock
{\BBOQ}\APACrefatitle {{The competition of electron and ion heating during magnetic reconnection}} {{The competition of electron and ion heating during magnetic reconnection}}.{\BBCQ}
\newblock
\APACjournalVolNumPages{Geophysical Research Letters}{42}{22}{9657-9665}.
\newblock
\begin{APACrefDOI} \doi{10.1002/2015GL065961} \end{APACrefDOI}
\PrintBackRefs{\CurrentBib}

\bibitem [\protect \citeauthoryear {%
{Harris}%
\ \protect \BOthers {.}}{%
{Harris}%
\ \protect \BOthers {.}}{%
{\protect \APACyear {2020}}%
}]{%
numpypaper}
\APACinsertmetastar {%
numpypaper}%
\begin{APACrefauthors}%
{Harris}, C\BPBI R.%
, {Millman}, K\BPBI J.%
, {van der Walt}, S\BPBI J.%
, {Gommers}, R.%
, {Virtanen}, P.%
, {Cournapeau}, D.%
\BDBL {}{Oliphant}, T\BPBI E.%
\end{APACrefauthors}%
\unskip\
\newblock
\APACrefYearMonthDay{2020}{{\APACmonth{09}}}{}.
\newblock
{\BBOQ}\APACrefatitle {{Array programming with NumPy}} {{Array programming with NumPy}}.{\BBCQ}
\newblock
\APACjournalVolNumPages{Nature}{585}{7825}{357-362}.
\newblock
\begin{APACrefDOI} \doi{10.1038/s41586-020-2649-2} \end{APACrefDOI}
\PrintBackRefs{\CurrentBib}

\bibitem [\protect \citeauthoryear {%
{Hoshino}%
}{%
{Hoshino}%
}{%
{\protect \APACyear {2023}}%
}]{%
hoshino2023apj}
\APACinsertmetastar {%
hoshino2023apj}%
\begin{APACrefauthors}%
{Hoshino}, M.%
\end{APACrefauthors}%
\unskip\
\newblock
\APACrefYearMonthDay{2023}{{\APACmonth{04}}}{}.
\newblock
{\BBOQ}\APACrefatitle {{Energy Partition of Thermal and Nonthermal Particles in Magnetic Reconnection}} {{Energy Partition of Thermal and Nonthermal Particles in Magnetic Reconnection}}.{\BBCQ}
\newblock
\APACjournalVolNumPages{Astrophysical Journal}{946}{2}{77}.
\newblock
\begin{APACrefDOI} \doi{10.3847/1538-4357/acbfb5} \end{APACrefDOI}
\PrintBackRefs{\CurrentBib}

\bibitem [\protect \citeauthoryear {%
{Huang}%
\ \protect \BOthers {.}}{%
{Huang}%
\ \protect \BOthers {.}}{%
{\protect \APACyear {2018}}%
}]{%
huang2018}
\APACinsertmetastar {%
huang2018}%
\begin{APACrefauthors}%
{Huang}, S\BPBI Y.%
, {Jiang}, K.%
, {Yuan}, Z\BPBI G.%
, {Sahraoui}, F.%
, {He}, L\BPBI H.%
, {Zhou}, M.%
\BDBL {}{Torbert}, R\BPBI B.%
\end{APACrefauthors}%
\unskip\
\newblock
\APACrefYearMonthDay{2018}{{\APACmonth{08}}}{}.
\newblock
{\BBOQ}\APACrefatitle {{Observations of the Electron Jet Generated by Secondary Reconnection in the Terrestrial Magnetotail}} {{Observations of the Electron Jet Generated by Secondary Reconnection in the Terrestrial Magnetotail}}.{\BBCQ}
\newblock
\APACjournalVolNumPages{Astrophysical Journal}{862}{2}{144}.
\newblock
\begin{APACrefDOI} \doi{10.3847/1538-4357/aacd4c} \end{APACrefDOI}
\PrintBackRefs{\CurrentBib}

\bibitem [\protect \citeauthoryear {%
{Imada}%
, {Hirai}%
\BCBL {}\ \BBA {} {Hoshino}%
}{%
{Imada}%
\ \protect \BOthers {.}}{%
{\protect \APACyear {2015}}%
}]{%
imada2015}
\APACinsertmetastar {%
imada2015}%
\begin{APACrefauthors}%
{Imada}, S.%
, {Hirai}, M.%
\BCBL {}\ \BBA {} {Hoshino}, M.%
\end{APACrefauthors}%
\unskip\
\newblock
\APACrefYearMonthDay{2015}{{\APACmonth{12}}}{}.
\newblock
{\BBOQ}\APACrefatitle {{Energetic ion acceleration during magnetic reconnection in the Earth's magnetotail}} {{Energetic ion acceleration during magnetic reconnection in the Earth's magnetotail}}.{\BBCQ}
\newblock
\APACjournalVolNumPages{Earth, Planets and Space}{67}{}{203}.
\newblock
\begin{APACrefDOI} \doi{10.1186/s40623-015-0372-2} \end{APACrefDOI}
\PrintBackRefs{\CurrentBib}

\bibitem [\protect \citeauthoryear {%
{Imada}%
, {Hirai}%
, {Hoshino}%
\BCBL {}\ \BBA {} {Mukai}%
}{%
{Imada}%
\ \protect \BOthers {.}}{%
{\protect \APACyear {2011}}%
}]{%
imada2011JGRA}
\APACinsertmetastar {%
imada2011JGRA}%
\begin{APACrefauthors}%
{Imada}, S.%
, {Hirai}, M.%
, {Hoshino}, M.%
\BCBL {}\ \BBA {} {Mukai}, T.%
\end{APACrefauthors}%
\unskip\
\newblock
\APACrefYearMonthDay{2011}{{\APACmonth{08}}}{}.
\newblock
{\BBOQ}\APACrefatitle {{Favorable conditions for energetic electron acceleration during magnetic reconnection in the Earth's magnetotail}} {{Favorable conditions for energetic electron acceleration during magnetic reconnection in the Earth's magnetotail}}.{\BBCQ}
\newblock
\APACjournalVolNumPages{Journal of Geophysical Research (Space Physics)}{116}{A8}{A08217}.
\newblock
\begin{APACrefDOI} \doi{10.1029/2011JA016576} \end{APACrefDOI}
\PrintBackRefs{\CurrentBib}

\bibitem [\protect \citeauthoryear {%
{Lavraud}%
\ \protect \BOthers {.}}{%
{Lavraud}%
\ \protect \BOthers {.}}{%
{\protect \APACyear {2009}}%
}]{%
lavraudgrl2009}
\APACinsertmetastar {%
lavraudgrl2009}%
\begin{APACrefauthors}%
{Lavraud}, B.%
, {Borovsky}, J\BPBI E.%
, {G{\'e}not}, V.%
, {Schwartz}, S\BPBI J.%
, {Birn}, J.%
, {Fazakerley}, A\BPBI N.%
\BDBL {}{Wild}, J\BPBI A.%
\end{APACrefauthors}%
\unskip\
\newblock
\APACrefYearMonthDay{2009}{{\APACmonth{09}}}{}.
\newblock
{\BBOQ}\APACrefatitle {{Tracing solar wind plasma entry into the magnetosphere using ion-to-electron temperature ratio}} {{Tracing solar wind plasma entry into the magnetosphere using ion-to-electron temperature ratio}}.{\BBCQ}
\newblock
\APACjournalVolNumPages{Geophysical Research Letters}{36}{18}{L18109}.
\newblock
\begin{APACrefDOI} \doi{10.1029/2009GL039442} \end{APACrefDOI}
\PrintBackRefs{\CurrentBib}

\bibitem [\protect \citeauthoryear {%
{Mauk}%
\ \protect \BOthers {.}}{%
{Mauk}%
\ \protect \BOthers {.}}{%
{\protect \APACyear {2016}}%
}]{%
epdeisfeepsmauk}
\APACinsertmetastar {%
epdeisfeepsmauk}%
\begin{APACrefauthors}%
{Mauk}, B\BPBI H.%
, {Blake}, J\BPBI B.%
, {Baker}, D\BPBI N.%
, {Clemmons}, J\BPBI H.%
, {Reeves}, G\BPBI D.%
, {Spence}, H\BPBI E.%
\BDBL {}{Westlake}, J\BPBI H.%
\end{APACrefauthors}%
\unskip\
\newblock
\APACrefYearMonthDay{2016}{{\APACmonth{03}}}{}.
\newblock
{\BBOQ}\APACrefatitle {{The Energetic Particle Detector (EPD) Investigation and the Energetic Ion Spectrometer (EIS) for the Magnetospheric Multiscale (MMS) Mission}} {{The Energetic Particle Detector (EPD) Investigation and the Energetic Ion Spectrometer (EIS) for the Magnetospheric Multiscale (MMS) Mission}}.{\BBCQ}
\newblock
\APACjournalVolNumPages{Space Science Reviews}{199}{1-4}{471-514}.
\newblock
\begin{APACrefDOI} \doi{10.1007/s11214-014-0055-5} \end{APACrefDOI}
\PrintBackRefs{\CurrentBib}

\bibitem [\protect \citeauthoryear {%
{Nagai}%
\ \BBA {} {Machida}%
}{%
{Nagai}%
\ \BBA {} {Machida}%
}{%
{\protect \APACyear {1998}}%
}]{%
nagai1998}
\APACinsertmetastar {%
nagai1998}%
\begin{APACrefauthors}%
{Nagai}, T.%
\BCBT {}\ \BBA {} {Machida}, S.%
\end{APACrefauthors}%
\unskip\
\newblock
\APACrefYearMonthDay{1998}{{\APACmonth{01}}}{}.
\newblock
{\BBOQ}\APACrefatitle {{Magnetic Reconnection in the Near-Earth Magnetotail}} {{Magnetic Reconnection in the Near-Earth Magnetotail}}.{\BBCQ}
\newblock
\APACjournalVolNumPages{Geophysical Monograph Series}{105}{}{211}.
\newblock
\begin{APACrefDOI} \doi{10.1029/GM105p0211} \end{APACrefDOI}
\PrintBackRefs{\CurrentBib}

\bibitem [\protect \citeauthoryear {%
{Nagai}%
, {Shinohara}%
\BCBL {}\ \BBA {} {Zenitani}%
}{%
{Nagai}%
\ \protect \BOthers {.}}{%
{\protect \APACyear {2015}}%
}]{%
nagai2015}
\APACinsertmetastar {%
nagai2015}%
\begin{APACrefauthors}%
{Nagai}, T.%
, {Shinohara}, I.%
\BCBL {}\ \BBA {} {Zenitani}, S.%
\end{APACrefauthors}%
\unskip\
\newblock
\APACrefYearMonthDay{2015}{{\APACmonth{03}}}{}.
\newblock
{\BBOQ}\APACrefatitle {{Ion acceleration processes in magnetic reconnection: Geotail observations in the magnetotail}} {{Ion acceleration processes in magnetic reconnection: Geotail observations in the magnetotail}}.{\BBCQ}
\newblock
\APACjournalVolNumPages{Journal of Geophysical Research (Space Physics)}{120}{3}{1766-1783}.
\newblock
\begin{APACrefDOI} \doi{10.1002/2014JA020737} \end{APACrefDOI}
\PrintBackRefs{\CurrentBib}

\bibitem [\protect \citeauthoryear {%
R.~{Nakamura}%
\ \protect \BOthers {.}}{%
R.~{Nakamura}%
\ \protect \BOthers {.}}{%
{\protect \APACyear {2002}}%
}]{%
nakamura2002}
\APACinsertmetastar {%
nakamura2002}%
\begin{APACrefauthors}%
{Nakamura}, R.%
, {Baumjohann}, W.%
, {Klecker}, B.%
, {Bogdanova}, Y.%
, {Balogh}, A.%
, {R{\`e}me}, H.%
\BDBL {}{Runov}, A.%
\end{APACrefauthors}%
\unskip\
\newblock
\APACrefYearMonthDay{2002}{{\APACmonth{10}}}{}.
\newblock
{\BBOQ}\APACrefatitle {{Motion of the dipolarization front during a flow burst event observed by Cluster}} {{Motion of the dipolarization front during a flow burst event observed by Cluster}}.{\BBCQ}
\newblock
\APACjournalVolNumPages{Geophysical Research Letters}{29}{20}{1942}.
\newblock
\begin{APACrefDOI} \doi{10.1029/2002GL015763} \end{APACrefDOI}
\PrintBackRefs{\CurrentBib}

\bibitem [\protect \citeauthoryear {%
R.~{Nakamura}%
\ \protect \BOthers {.}}{%
R.~{Nakamura}%
\ \protect \BOthers {.}}{%
{\protect \APACyear {2019}}%
}]{%
nakamura2019jgr}
\APACinsertmetastar {%
nakamura2019jgr}%
\begin{APACrefauthors}%
{Nakamura}, R.%
, {Genestreti}, K\BPBI J.%
, {Nakamura}, T.%
, {Baumjohann}, W.%
, {Varsani}, A.%
, {Nagai}, T.%
\BDBL {}{Torbert}, R\BPBI B.%
\end{APACrefauthors}%
\unskip\
\newblock
\APACrefYearMonthDay{2019}{{\APACmonth{02}}}{}.
\newblock
{\BBOQ}\APACrefatitle {{Structure of the Current Sheet in the 11 July 2017 Electron Diffusion Region Event}} {{Structure of the Current Sheet in the 11 July 2017 Electron Diffusion Region Event}}.{\BBCQ}
\newblock
\APACjournalVolNumPages{Journal of Geophysical Research (Space Physics)}{124}{2}{1173-1186}.
\newblock
\begin{APACrefDOI} \doi{10.1029/2018JA026028} \end{APACrefDOI}
\PrintBackRefs{\CurrentBib}

\bibitem [\protect \citeauthoryear {%
T\BPBI K\BPBI M.~{Nakamura}%
\ \protect \BOthers {.}}{%
T\BPBI K\BPBI M.~{Nakamura}%
\ \protect \BOthers {.}}{%
{\protect \APACyear {2018}}%
}]{%
nakamura2018jgr}
\APACinsertmetastar {%
nakamura2018jgr}%
\begin{APACrefauthors}%
{Nakamura}, T\BPBI K\BPBI M.%
, {Genestreti}, K\BPBI J.%
, {Liu}, Y\BPBI H.%
, {Nakamura}, R.%
, {Teh}, W\BPBI L.%
, {Hasegawa}, H.%
\BDBL {}{Giles}, B\BPBI L.%
\end{APACrefauthors}%
\unskip\
\newblock
\APACrefYearMonthDay{2018}{{\APACmonth{11}}}{}.
\newblock
{\BBOQ}\APACrefatitle {{Measurement of the Magnetic Reconnection Rate in the Earth's Magnetotail}} {{Measurement of the Magnetic Reconnection Rate in the Earth's Magnetotail}}.{\BBCQ}
\newblock
\APACjournalVolNumPages{Journal of Geophysical Research (Space Physics)}{123}{11}{9150-9168}.
\newblock
\begin{APACrefDOI} \doi{10.1029/2018JA025713} \end{APACrefDOI}
\PrintBackRefs{\CurrentBib}

\bibitem [\protect \citeauthoryear {%
{{\O}ieroset}%
, {Lin}%
, {Phan}%
, {Larson}%
\BCBL {}\ \BBA {} {Bale}%
}{%
{{\O}ieroset}%
\ \protect \BOthers {.}}{%
{\protect \APACyear {2002}}%
}]{%
oieroset2002}
\APACinsertmetastar {%
oieroset2002}%
\begin{APACrefauthors}%
{{\O}ieroset}, M.%
, {Lin}, R\BPBI P.%
, {Phan}, T\BPBI D.%
, {Larson}, D\BPBI E.%
\BCBL {}\ \BBA {} {Bale}, S\BPBI D.%
\end{APACrefauthors}%
\unskip\
\newblock
\APACrefYearMonthDay{2002}{{\APACmonth{10}}}{}.
\newblock
{\BBOQ}\APACrefatitle {{Evidence for Electron Acceleration up to \raisebox{-0.5ex}\textasciitilde300 keV in the Magnetic Reconnection Diffusion Region of Earth's Magnetotail}} {{Evidence for Electron Acceleration up to \raisebox{-0.5ex}\textasciitilde300 keV in the Magnetic Reconnection Diffusion Region of Earth's Magnetotail}}.{\BBCQ}
\newblock
\APACjournalVolNumPages{Physical Review Letters}{89}{19}{195001}.
\newblock
\begin{APACrefDOI} \doi{10.1103/PhysRevLett.89.195001} \end{APACrefDOI}
\PrintBackRefs{\CurrentBib}

\bibitem [\protect \citeauthoryear {%
{{\O}ieroset}%
\ \protect \BOthers {.}}{%
{{\O}ieroset}%
\ \protect \BOthers {.}}{%
{\protect \APACyear {2023}}%
}]{%
oierioset2023}
\APACinsertmetastar {%
oierioset2023}%
\begin{APACrefauthors}%
{{\O}ieroset}, M.%
, {Phan}, T\BPBI D.%
, {Oka}, M.%
, {Drake}, J\BPBI F.%
, {Fuselier}, S\BPBI A.%
, {Gershman}, D\BPBI J.%
\BDBL {}{Strangeway}, R\BPBI J.%
\end{APACrefauthors}%
\unskip\
\newblock
\APACrefYearMonthDay{2023}{{\APACmonth{09}}}{}.
\newblock
{\BBOQ}\APACrefatitle {{Scaling of Magnetic Reconnection Electron Bulk Heating in the High-Alfv{\'e}n-speed and Low-{\ensuremath{\beta}} Regime of Earth's Magnetotail}} {{Scaling of Magnetic Reconnection Electron Bulk Heating in the High-Alfv{\'e}n-speed and Low-{\ensuremath{\beta}} Regime of Earth's Magnetotail}}.{\BBCQ}
\newblock
\APACjournalVolNumPages{Astrophysical Journal}{954}{2}{118}.
\newblock
\begin{APACrefDOI} \doi{10.3847/1538-4357/acdf44} \end{APACrefDOI}
\PrintBackRefs{\CurrentBib}

\bibitem [\protect \citeauthoryear {%
{Oka}%
, {Ishikawa}%
, {Saint-Hilaire}%
, {Krucker}%
\BCBL {}\ \BBA {} {Lin}%
}{%
{Oka}%
\ \protect \BOthers {.}}{%
{\protect \APACyear {2013}}%
}]{%
oka2013}
\APACinsertmetastar {%
oka2013}%
\begin{APACrefauthors}%
{Oka}, M.%
, {Ishikawa}, S.%
, {Saint-Hilaire}, P.%
, {Krucker}, S.%
\BCBL {}\ \BBA {} {Lin}, R\BPBI P.%
\end{APACrefauthors}%
\unskip\
\newblock
\APACrefYearMonthDay{2013}{{\APACmonth{02}}}{}.
\newblock
{\BBOQ}\APACrefatitle {{Kappa Distribution Model for Hard X-Ray Coronal Sources of Solar Flares}} {{Kappa Distribution Model for Hard X-Ray Coronal Sources of Solar Flares}}.{\BBCQ}
\newblock
\APACjournalVolNumPages{Astrophysical Journal}{764}{1}{6}.
\newblock
\begin{APACrefDOI} \doi{10.1088/0004-637X/764/1/6} \end{APACrefDOI}
\PrintBackRefs{\CurrentBib}

\bibitem [\protect \citeauthoryear {%
{Oka}%
, {Krucker}%
, {Hudson}%
\BCBL {}\ \BBA {} {Saint-Hilaire}%
}{%
{Oka}%
\ \protect \BOthers {.}}{%
{\protect \APACyear {2015}}%
}]{%
oka2015}
\APACinsertmetastar {%
oka2015}%
\begin{APACrefauthors}%
{Oka}, M.%
, {Krucker}, S.%
, {Hudson}, H\BPBI S.%
\BCBL {}\ \BBA {} {Saint-Hilaire}, P.%
\end{APACrefauthors}%
\unskip\
\newblock
\APACrefYearMonthDay{2015}{{\APACmonth{02}}}{}.
\newblock
{\BBOQ}\APACrefatitle {{Electron Energy Partition in the Above-the-looptop Solar Hard X-Ray Sources}} {{Electron Energy Partition in the Above-the-looptop Solar Hard X-Ray Sources}}.{\BBCQ}
\newblock
\APACjournalVolNumPages{Astrophysical Journal}{799}{2}{129}.
\newblock
\begin{APACrefDOI} \doi{10.1088/0004-637X/799/2/129} \end{APACrefDOI}
\PrintBackRefs{\CurrentBib}

\bibitem [\protect \citeauthoryear {%
{Oka}%
\ \protect \BOthers {.}}{%
{Oka}%
\ \protect \BOthers {.}}{%
{\protect \APACyear {2022}}%
}]{%
okaetal2022}
\APACinsertmetastar {%
okaetal2022}%
\begin{APACrefauthors}%
{Oka}, M.%
, {Phan}, T.%
, {{\O}ieroset}, M.%
, {Turner}, D.%
, {Drake}, J.%
, {Li}, X.%
\BDBL {}{Burch}, J.%
\end{APACrefauthors}%
\unskip\
\newblock
\APACrefYearMonthDay{2022}{{\APACmonth{05}}}{}.
\newblock
{\BBOQ}\APACrefatitle {{Electron energization and thermal to non- thermal energy partition during earth's magnetotail reconnection}} {{Electron energization and thermal to non- thermal energy partition during earth's magnetotail reconnection}}.{\BBCQ}
\newblock
\APACjournalVolNumPages{Physics of Plasmas}{29}{5}{052904}.
\newblock
\begin{APACrefDOI} \doi{10.1063/5.0085647} \end{APACrefDOI}
\PrintBackRefs{\CurrentBib}

\bibitem [\protect \citeauthoryear {%
{Olbert}%
}{%
{Olbert}%
}{%
{\protect \APACyear {1968}}%
}]{%
olbert1968}
\APACinsertmetastar {%
olbert1968}%
\begin{APACrefauthors}%
{Olbert}, S.%
\end{APACrefauthors}%
\unskip\
\newblock
\APACrefYearMonthDay{1968}{{\APACmonth{01}}}{}.
\newblock
{\BBOQ}\APACrefatitle {{Summary of Experimental Results from M.I.T. Detector on IMP-1}} {{Summary of Experimental Results from M.I.T. Detector on IMP-1}}.{\BBCQ}
\newblock
\BIn{} R\BPBI D\BPBI L.~{Carovillano}\ \BBA {} J\BPBI F.~{McClay}\ (\BEDS), \APACrefbtitle {Physics of the Magnetosphere} {Physics of the magnetosphere}\ (\BVOL~10, \BPG~641).
\newblock
\begin{APACrefDOI} \doi{10.1007/978-94-010-3467-8_23} \end{APACrefDOI}
\PrintBackRefs{\CurrentBib}

\bibitem [\protect \citeauthoryear {%
{Pierrard}%
\ \BBA {} {Lazar}%
}{%
{Pierrard}%
\ \BBA {} {Lazar}%
}{%
{\protect \APACyear {2010}}%
}]{%
pierrard2010}
\APACinsertmetastar {%
pierrard2010}%
\begin{APACrefauthors}%
{Pierrard}, V.%
\BCBT {}\ \BBA {} {Lazar}, M.%
\end{APACrefauthors}%
\unskip\
\newblock
\APACrefYearMonthDay{2010}{{\APACmonth{11}}}{}.
\newblock
{\BBOQ}\APACrefatitle {{Kappa Distributions: Theory and Applications in Space Plasmas}} {{Kappa Distributions: Theory and Applications in Space Plasmas}}.{\BBCQ}
\newblock
\APACjournalVolNumPages{Solar Physics}{267}{1}{153-174}.
\newblock
\begin{APACrefDOI} \doi{10.1007/s11207-010-9640-2} \end{APACrefDOI}
\PrintBackRefs{\CurrentBib}

\bibitem [\protect \citeauthoryear {%
{Pollock}%
\ \protect \BOthers {.}}{%
{Pollock}%
\ \protect \BOthers {.}}{%
{\protect \APACyear {2016}}%
}]{%
fpipollock}
\APACinsertmetastar {%
fpipollock}%
\begin{APACrefauthors}%
{Pollock}, C.%
, {Moore}, T.%
, {Jacques}, A.%
, {Burch}, J.%
, {Gliese}, U.%
, {Saito}, Y.%
\BDBL {}{Zeuch}, M.%
\end{APACrefauthors}%
\unskip\
\newblock
\APACrefYearMonthDay{2016}{{\APACmonth{03}}}{}.
\newblock
{\BBOQ}\APACrefatitle {{Fast Plasma Investigation for Magnetospheric Multiscale}} {{Fast Plasma Investigation for Magnetospheric Multiscale}}.{\BBCQ}
\newblock
\APACjournalVolNumPages{Space Science Reviews}{199}{1-4}{331-406}.
\newblock
\begin{APACrefDOI} \doi{10.1007/s11214-016-0245-4} \end{APACrefDOI}
\PrintBackRefs{\CurrentBib}

\bibitem [\protect \citeauthoryear {%
{Priest}%
\ \BBA {} {Forbes}%
}{%
{Priest}%
\ \BBA {} {Forbes}%
}{%
{\protect \APACyear {2000}}%
}]{%
priestforbes}
\APACinsertmetastar {%
priestforbes}%
\begin{APACrefauthors}%
{Priest}, E.%
\BCBT {}\ \BBA {} {Forbes}, T.%
\end{APACrefauthors}%
\unskip\
\newblock
\APACrefYear{2000}.
\newblock
\APACrefbtitle {{Magnetic Reconnection: MHD Theory and Applications}} {{Magnetic Reconnection: MHD Theory and Applications}}.
\newblock
\begin{APACrefDOI} \doi{10.1017/CBO9780511525087} \end{APACrefDOI}
\PrintBackRefs{\CurrentBib}

\bibitem [\protect \citeauthoryear {%
{Rogers}%
, {Farrugia}%
\BCBL {}\ \BBA {} {Torbert}%
}{%
{Rogers}%
\ \protect \BOthers {.}}{%
{\protect \APACyear {2019}}%
}]{%
rogers2019}
\APACinsertmetastar {%
rogers2019}%
\begin{APACrefauthors}%
{Rogers}, A\BPBI J.%
, {Farrugia}, C\BPBI J.%
\BCBL {}\ \BBA {} {Torbert}, R\BPBI B.%
\end{APACrefauthors}%
\unskip\
\newblock
\APACrefYearMonthDay{2019}{{\APACmonth{08}}}{}.
\newblock
{\BBOQ}\APACrefatitle {{Numerical Algorithm for Detecting Ion Diffusion Regions in the Geomagnetic Tail With Applications to MMS Tail Season 1 May to 30 September 2017}} {{Numerical Algorithm for Detecting Ion Diffusion Regions in the Geomagnetic Tail With Applications to MMS Tail Season 1 May to 30 September 2017}}.{\BBCQ}
\newblock
\APACjournalVolNumPages{Journal of Geophysical Research (Space Physics)}{124}{8}{6487-6503}.
\newblock
\begin{APACrefDOI} \doi{10.1029/2018JA026429} \end{APACrefDOI}
\PrintBackRefs{\CurrentBib}

\bibitem [\protect \citeauthoryear {%
{Slavin}%
\ \protect \BOthers {.}}{%
{Slavin}%
\ \protect \BOthers {.}}{%
{\protect \APACyear {1985}}%
}]{%
slavin1985jgr}
\APACinsertmetastar {%
slavin1985jgr}%
\begin{APACrefauthors}%
{Slavin}, J\BPBI A.%
, {Smith}, E\BPBI J.%
, {Sibeck}, D\BPBI G.%
, {Baker}, D\BPBI N.%
, {Zwickl}, R\BPBI D.%
\BCBL {}\ \BBA {} {Akasofu}, S\BPBI I.%
\end{APACrefauthors}%
\unskip\
\newblock
\APACrefYearMonthDay{1985}{{\APACmonth{11}}}{}.
\newblock
{\BBOQ}\APACrefatitle {{An ISEE 3 study of average and substorm conditions in the distant magnetotail}} {{An ISEE 3 study of average and substorm conditions in the distant magnetotail}}.{\BBCQ}
\newblock
\APACjournalVolNumPages{Journal of Geophysical Research}{90}{A11}{10875-10895}.
\newblock
\begin{APACrefDOI} \doi{10.1029/JA090iA11p10875} \end{APACrefDOI}
\PrintBackRefs{\CurrentBib}

\bibitem [\protect \citeauthoryear {%
{Torbert}%
\ \protect \BOthers {.}}{%
{Torbert}%
\ \protect \BOthers {.}}{%
{\protect \APACyear {2018}}%
}]{%
torbert2018science}
\APACinsertmetastar {%
torbert2018science}%
\begin{APACrefauthors}%
{Torbert}, R\BPBI B.%
, {Burch}, J\BPBI L.%
, {Phan}, T\BPBI D.%
, {Hesse}, M.%
, {Argall}, M\BPBI R.%
, {Shuster}, J.%
\BDBL {}{Saito}, Y.%
\end{APACrefauthors}%
\unskip\
\newblock
\APACrefYearMonthDay{2018}{{\APACmonth{12}}}{}.
\newblock
{\BBOQ}\APACrefatitle {{Electron-scale dynamics of the diffusion region during symmetric magnetic reconnection in space}} {{Electron-scale dynamics of the diffusion region during symmetric magnetic reconnection in space}}.{\BBCQ}
\newblock
\APACjournalVolNumPages{Science}{362}{6421}{1391-1395}.
\newblock
\begin{APACrefDOI} \doi{10.1126/science.aat2998} \end{APACrefDOI}
\PrintBackRefs{\CurrentBib}

\bibitem [\protect \citeauthoryear {%
{Torbert}%
\ \protect \BOthers {.}}{%
{Torbert}%
\ \protect \BOthers {.}}{%
{\protect \APACyear {2016}}%
}]{%
fieldstorbert}
\APACinsertmetastar {%
fieldstorbert}%
\begin{APACrefauthors}%
{Torbert}, R\BPBI B.%
, {Russell}, C\BPBI T.%
, {Magnes}, W.%
, {Ergun}, R\BPBI E.%
, {Lindqvist}, P\BPBI A.%
, {Le Contel}, O.%
\BDBL {}{Lappalainen}, K.%
\end{APACrefauthors}%
\unskip\
\newblock
\APACrefYearMonthDay{2016}{{\APACmonth{03}}}{}.
\newblock
{\BBOQ}\APACrefatitle {{The FIELDS Instrument Suite on MMS: Scientific Objectives, Measurements, and Data Products}} {{The FIELDS Instrument Suite on MMS: Scientific Objectives, Measurements, and Data Products}}.{\BBCQ}
\newblock
\APACjournalVolNumPages{Space Science Reviews}{199}{1-4}{105-135}.
\newblock
\begin{APACrefDOI} \doi{10.1007/s11214-014-0109-8} \end{APACrefDOI}
\PrintBackRefs{\CurrentBib}

\bibitem [\protect \citeauthoryear {%
Virtanen%
\ \protect \BOthers {.}}{%
Virtanen%
\ \protect \BOthers {.}}{%
{\protect \APACyear {2020}}%
}]{%
scipypaper}
\APACinsertmetastar {%
scipypaper}%
\begin{APACrefauthors}%
Virtanen, P.%
, Gommers, R.%
, Oliphant, T\BPBI E.%
, Haberland, M.%
, Reddy, T.%
, Cournapeau, D.%
\BDBL {}{SciPy 1.0 Contributors}%
\end{APACrefauthors}%
\unskip\
\newblock
\APACrefYearMonthDay{2020}{}{}.
\newblock
{\BBOQ}\APACrefatitle {{{SciPy} 1.0: Fundamental Algorithms for Scientific Computing in Python}} {{{SciPy} 1.0: Fundamental Algorithms for Scientific Computing in Python}}.{\BBCQ}
\newblock
\APACjournalVolNumPages{Nature Methods}{17}{}{261--272}.
\newblock
\begin{APACrefDOI} \doi{10.1038/s41592-019-0686-2} \end{APACrefDOI}
\PrintBackRefs{\CurrentBib}

\bibitem [\protect \citeauthoryear {%
{Yin}%
, {Drake}%
\BCBL {}\ \BBA {} {Swisdak}%
}{%
{Yin}%
\ \protect \BOthers {.}}{%
{\protect \APACyear {2024}}%
}]{%
zhiyuyin2024}
\APACinsertmetastar {%
zhiyuyin2024}%
\begin{APACrefauthors}%
{Yin}, Z.%
, {Drake}, J\BPBI F.%
\BCBL {}\ \BBA {} {Swisdak}, M.%
\end{APACrefauthors}%
\unskip\
\newblock
\APACrefYearMonthDay{2024}{{\APACmonth{10}}}{}.
\newblock
{\BBOQ}\APACrefatitle {{Simultaneous Proton and Electron Energization during Macroscale Magnetic Reconnection}} {{Simultaneous Proton and Electron Energization during Macroscale Magnetic Reconnection}}.{\BBCQ}
\newblock
\APACjournalVolNumPages{Astrophysical Journal}{974}{1}{74}.
\newblock
\begin{APACrefDOI} \doi{10.3847/1538-4357/ad7131} \end{APACrefDOI}
\PrintBackRefs{\CurrentBib}

\bibitem [\protect \citeauthoryear {%
{Young}%
\ \protect \BOthers {.}}{%
{Young}%
\ \protect \BOthers {.}}{%
{\protect \APACyear {2016}}%
}]{%
hpcayoung}
\APACinsertmetastar {%
hpcayoung}%
\begin{APACrefauthors}%
{Young}, D\BPBI T.%
, {Burch}, J\BPBI L.%
, {Gomez}, R\BPBI G.%
, {De Los Santos}, A.%
, {Miller}, G\BPBI P.%
, {Wilson}, P.%
\BDBL {}{Webster}, J\BPBI M.%
\end{APACrefauthors}%
\unskip\
\newblock
\APACrefYearMonthDay{2016}{{\APACmonth{03}}}{}.
\newblock
{\BBOQ}\APACrefatitle {{Hot Plasma Composition Analyzer for the Magnetospheric Multiscale Mission}} {{Hot Plasma Composition Analyzer for the Magnetospheric Multiscale Mission}}.{\BBCQ}
\newblock
\APACjournalVolNumPages{Space Science Reviews}{199}{1-4}{407-470}.
\newblock
\begin{APACrefDOI} \doi{10.1007/s11214-014-0119-6} \end{APACrefDOI}
\PrintBackRefs{\CurrentBib}

\bibitem [\protect \citeauthoryear {%
{Zhang}%
, {Guo}%
, {Daughton}%
, {Li}%
\BCBL {}\ \BBA {} {Li}%
}{%
{Zhang}%
\ \protect \BOthers {.}}{%
{\protect \APACyear {2021}}%
}]{%
zhang2021}
\APACinsertmetastar {%
zhang2021}%
\begin{APACrefauthors}%
{Zhang}, Q.%
, {Guo}, F.%
, {Daughton}, W.%
, {Li}, H.%
\BCBL {}\ \BBA {} {Li}, X.%
\end{APACrefauthors}%
\unskip\
\newblock
\APACrefYearMonthDay{2021}{{\APACmonth{10}}}{}.
\newblock
{\BBOQ}\APACrefatitle {{Efficient Nonthermal Ion and Electron Acceleration Enabled by the Flux-Rope Kink Instability in 3D Nonrelativistic Magnetic Reconnection}} {{Efficient Nonthermal Ion and Electron Acceleration Enabled by the Flux-Rope Kink Instability in 3D Nonrelativistic Magnetic Reconnection}}.{\BBCQ}
\newblock
\APACjournalVolNumPages{Physical Review Letters}{127}{18}{185101}.
\newblock
\begin{APACrefDOI} \doi{10.1103/PhysRevLett.127.185101} \end{APACrefDOI}
\PrintBackRefs{\CurrentBib}

\bibitem [\protect \citeauthoryear {%
{Zhou}%
\ \protect \BOthers {.}}{%
{Zhou}%
\ \protect \BOthers {.}}{%
{\protect \APACyear {2019}}%
}]{%
zhou2019}
\APACinsertmetastar {%
zhou2019}%
\begin{APACrefauthors}%
{Zhou}, M.%
, {Deng}, X\BPBI H.%
, {Zhong}, Z\BPBI H.%
, {Pang}, Y.%
, {Tang}, R\BPBI X.%
, {El-Alaoui}, M.%
\BDBL {}{Lindqvist}, P\BPBI A.%
\end{APACrefauthors}%
\unskip\
\newblock
\APACrefYearMonthDay{2019}{{\APACmonth{01}}}{}.
\newblock
{\BBOQ}\APACrefatitle {{Observations of an Electron Diffusion Region in Symmetric Reconnection with Weak Guide Field}} {{Observations of an Electron Diffusion Region in Symmetric Reconnection with Weak Guide Field}}.{\BBCQ}
\newblock
\APACjournalVolNumPages{Astrophysical Journal}{870}{1}{34}.
\newblock
\begin{APACrefDOI} \doi{10.3847/1538-4357/aaf16f} \end{APACrefDOI}
\PrintBackRefs{\CurrentBib}

\bibitem [\protect \citeauthoryear {%
{Zhou}%
\ \protect \BOthers {.}}{%
{Zhou}%
\ \protect \BOthers {.}}{%
{\protect \APACyear {2016}}%
}]{%
zhou2016jgr}
\APACinsertmetastar {%
zhou2016jgr}%
\begin{APACrefauthors}%
{Zhou}, M.%
, {Li}, T.%
, {Deng}, X.%
, {Pang}, Y.%
, {Xu}, X.%
, {Tang}, R.%
\BDBL {}{Li}, H.%
\end{APACrefauthors}%
\unskip\
\newblock
\APACrefYearMonthDay{2016}{{\APACmonth{04}}}{}.
\newblock
{\BBOQ}\APACrefatitle {{Statistics of energetic electrons in the magnetotail reconnection}} {{Statistics of energetic electrons in the magnetotail reconnection}}.{\BBCQ}
\newblock
\APACjournalVolNumPages{Journal of Geophysical Research (Space Physics)}{121}{4}{3108-3119}.
\newblock
\begin{APACrefDOI} \doi{10.1002/2015JA022085} \end{APACrefDOI}
\PrintBackRefs{\CurrentBib}

\bibitem [\protect \citeauthoryear {%
{Zweibel}%
\ \BBA {} {Yamada}%
}{%
{Zweibel}%
\ \BBA {} {Yamada}%
}{%
{\protect \APACyear {2009}}%
}]{%
yamadareview}
\APACinsertmetastar {%
yamadareview}%
\begin{APACrefauthors}%
{Zweibel}, E\BPBI G.%
\BCBT {}\ \BBA {} {Yamada}, M.%
\end{APACrefauthors}%
\unskip\
\newblock
\APACrefYearMonthDay{2009}{{\APACmonth{09}}}{}.
\newblock
{\BBOQ}\APACrefatitle {{Magnetic Reconnection in Astrophysical and Laboratory Plasmas}} {{Magnetic Reconnection in Astrophysical and Laboratory Plasmas}}.{\BBCQ}
\newblock
\APACjournalVolNumPages{Annual Review of Astronomy and Astrophysics}{47}{1}{291-332}.
\newblock
\begin{APACrefDOI} \doi{10.1146/annurev-astro-082708-101726} \end{APACrefDOI}
\PrintBackRefs{\CurrentBib}

\end{thebibliography}
\end{document}